\documentclass[journal,twoside,web]{IEEEtran}
\usepackage{generic}
\usepackage{cite}
\usepackage{amsmath,amssymb,amsfonts}
\usepackage{algorithmic}
\usepackage{graphicx}
\usepackage{textcomp}
\def\BibTeX{{\rm B\kern-.05em{\sc i\kern-.025em b}\kern-.08em
    T\kern-.1667em\lower.7ex\hbox{E}\kern-.125emX}}

\usepackage{silence}
\usepackage{mathtools}
\usepackage{commath}
\usepackage{diagbox}
\usepackage{multirow}
\usepackage{hhline}
\usepackage{subfig}
\usepackage{float}
\usepackage{soul}
\usepackage{xcolor}
\usepackage[normalem]{ulem}
\usepackage{upgreek}
\usepackage{microtype}
\useunder{\uline}{\ul}{}
\usepackage{hyperref}

\usepackage{flushend} 

\DisableLigatures{encoding = *, family = * }

\newcommand{\rvect}[1]{\begin{bsmallmatrix} #1 \end{bsmallmatrix}}

\newcommand{\ts}{\textsuperscript}
\newcommand{\etal}{~\textit{et~al.}}

\def\NV#1{#1} 
\def\NVN#1{#1} 
\def\hl#1{#1}

\begin{document}

\title{Segmentation of Bruch's Membrane in retinal OCT with AMD using anatomical priors and uncertainty quantification}

\author{\authorblockN{Botond Fazekas, Dmitrii Lachinov, Guilherme Aresta, Julia Mai, Ursula Schmidt-Erfurth,\\ and Hrvoje Bogunovi\'c}\\
\authorblockA{Christian Doppler Laboratory for Artificial Intelligence in Retina, Department of Ophthalmology and Optometry, Medical University of Vienna, Austria}
}

\maketitle

\begin{abstract}
Bruch’s membrane (BM) segmentation on optical coherence tomography (OCT) is a pivotal step for the diagnosis and follow-up of age-related macular degeneration (AMD), one of the leading causes of blindness in the developed world. Automated BM segmentation methods exist, but they usually do not account for the anatomical coherence of the results, neither provide feedback on the confidence of the prediction. These factors limit the applicability of these systems in real-world scenarios. With this in mind, we propose an end-to-end deep learning method for automated BM segmentation in AMD patients. An Attention U-Net is trained to output a probability density function of the BM position, while taking into account the natural curvature of the surface. Besides the surface position, the method also estimates an A-scan wise uncertainty measure of the segmentation output. Subsequently, the A-scans with high uncertainty are interpolated using thin plate splines (TPS). We tested our method with ablation studies on an internal dataset with 138 patients covering all three AMD stages, and achieved a mean absolute localization error of 4.10~$\mathbf{\upmu}$m. In addition, the proposed segmentation method was compared against the state-of-the-art methods and showed a superior performance on an external publicly available dataset from a different patient cohort and OCT device, demonstrating strong generalization ability.
\end{abstract}

\section{Introduction}

Age-related Macular Degeneration (AMD) is the leading cause of blindness and irreversible loss of central vision in the developed countries for people over age sixty \cite{2004_Bressler}. The disease stages are divided into early/intermediate (iAMD) and two late ones. Early/intermediate stage AMD cases are characterized by the deposition of metabolic products (drusen) in the macula between the retinal pigment epithelium (RPE) and the Bruch's Membrane (BM). Late AMD appears as geographic atrophy (GA or dry AMD) or neovascular AMD (nAMD or wet AMD), although a mixture of both can occur in the same eye~\cite{2020_Stahl}. GA is characterized by the progressive loss of photoreceptors and RPE in the macular region, resulting in the permanent loss of the sharp vision. Neovascular AMD manifests in the formation of abnormal blood vessels, typically in the choroidal plexus below the RPE, leading to pigment epithelial detachment (PED), and exudation into the retina. Thus, early diagnosis and regular monitoring is crucial for the effective treatment of the patients. Although AMD is usually detected first by funduscopic examinations, other imaging modalities are required to understand the full extent of the degeneration under the macula \cite{2020_Stahl}. The current state-of-the-art modality for AMD monitoring and treatment is Optical Coherence Tomography.

Optical Coherence Tomography (OCT) is a non-invasive 3D imaging technique that can acquire high-resolution cross-sectional images of human tissues, particularly suitable for the retina \cite{1991_Huang,1995_Hee}. It is widely used in the diagnosis and monitoring of patients with a large variety of retinal diseases, such as diabetic retinopathy (DR) \cite{2016_Bavinger}, retinal vein occlusion (RVO) \cite{2016_Coscas}, glaucoma \cite{1995_Puliafito} and AMD \cite{2011_Malamos,2017_Fang}. The segmentation of the retinal layers in OCT scans is a crucial step in order to monitor and quantify the progression of a disease. However, manual layer annotation or correction is very time-consuming and subjective, which has motivated the development of automated accurate and objective methods \cite{2019_He_CONF, 2017_Roy, 2020_Orlando, 2021_Sousa}.

\begin{figure}[tb]
    \begin{minipage}{.15\textwidth}
    \centering
        \includegraphics[width=1.0\textwidth]{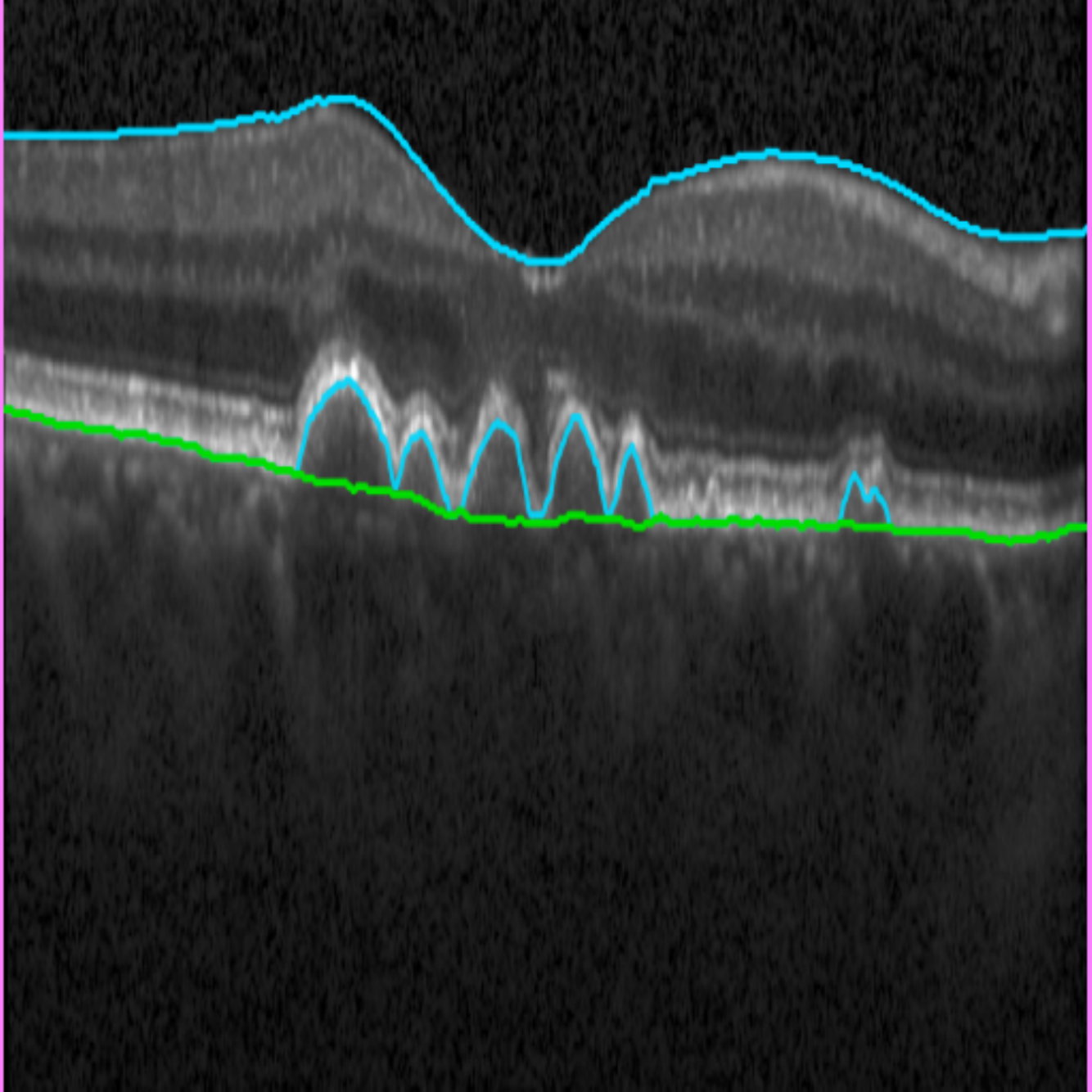}
        \caption*{(a) iAMD}
    \end{minipage}
    \hfill
    \begin{minipage}{.15\textwidth}
        \centering
        \includegraphics[width=1.0\textwidth]{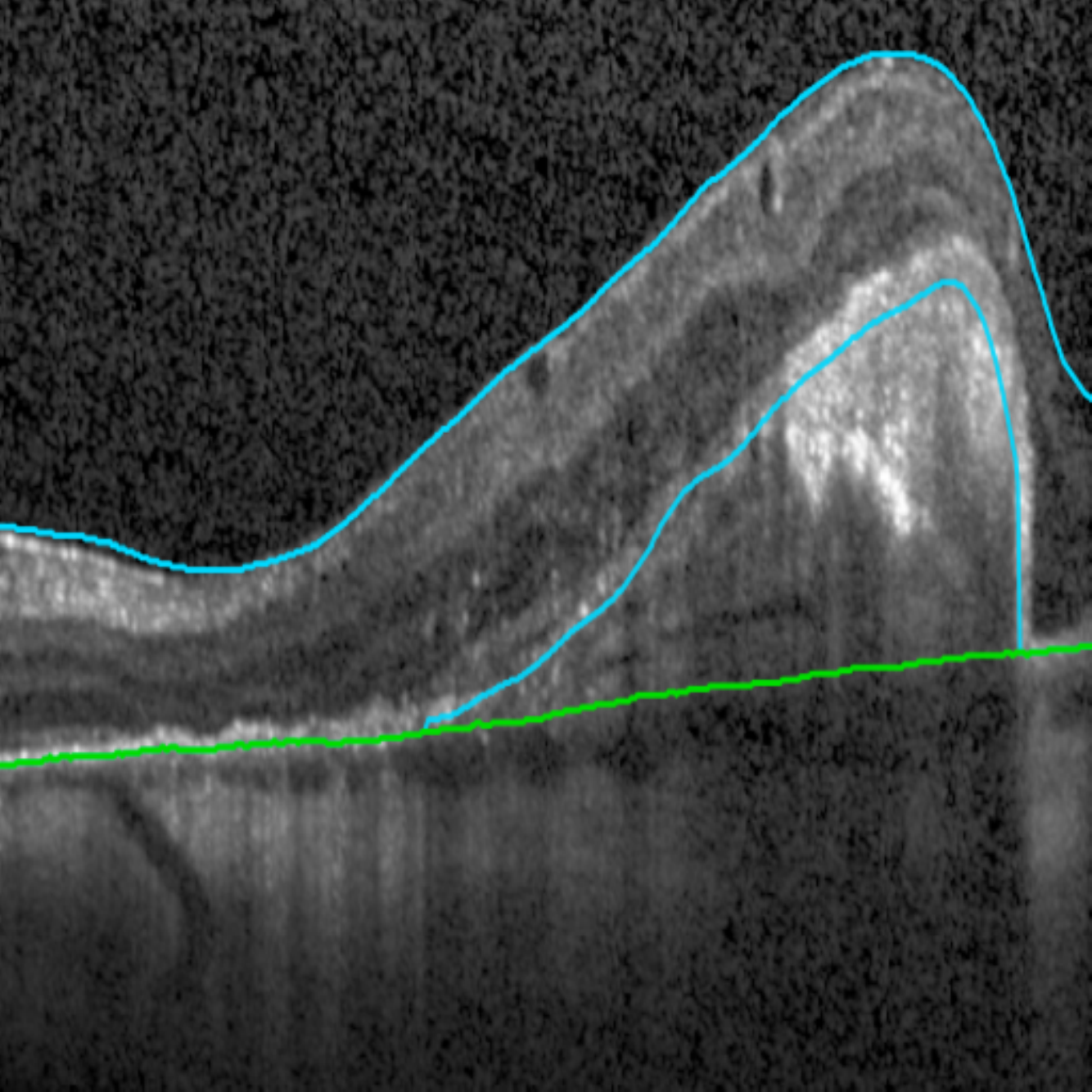}
        \caption*{(b) nAMD}
    \end{minipage}
    \hfill
    \begin{minipage}{.15\textwidth}
        \centering
        \includegraphics[width=1.0\textwidth]{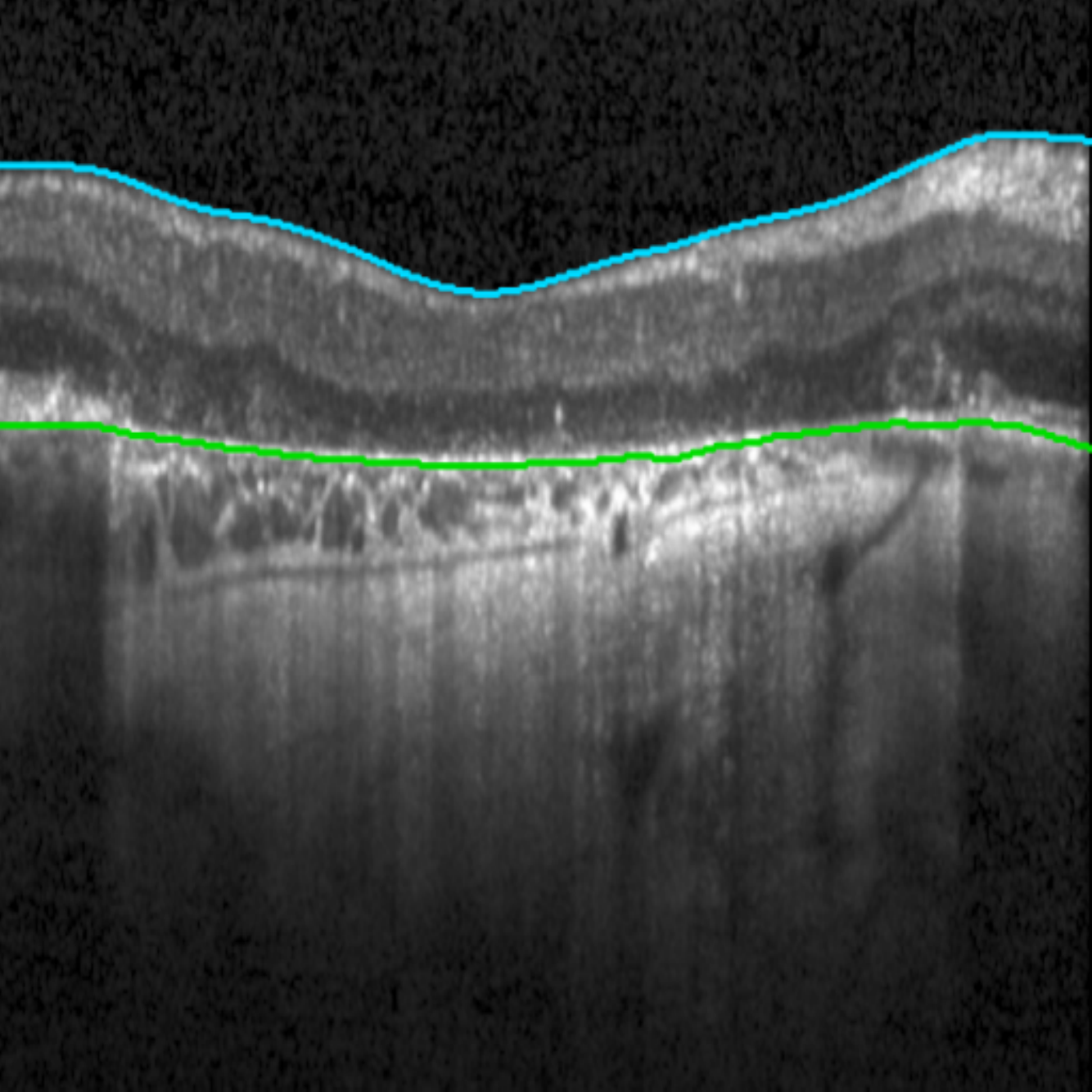}
        \caption*{(c) GA}
    \end{minipage}
    \caption{The three stages of AMD, where the Bruch's Membrane is marked with a green line. \NV{As a reference, retina from the Internal Limiting Membrane to the outer boundary of the RPE is denoted with cyan lines.}}
    \label{fig:amd_samples}
\end{figure}

Bruch’s Membrane is an elastic smooth and thin structure, strategically located between the retina and the general circulation, having a crucial role in retinal function, aging and disease~\hl{\cite{2010_Booij}}. Automated segmentation of the BM is particularly important in the context of AMD as, unlike other common retinal diseases such as DR, RVO, or glaucoma, the BM is distinguishable from the outer RPE boundary. In specific, drusen in iAMD and PEDs in nAMD separate the RPE from BM, requiring the segmentation of the region in-between them. In addition, in case of GA, the RPE is completely lost in some locations, exposing only the BM, thus imposing additional difficulties for algorithms and calculations that depend on the RPE position. Achieving correct automated identification of the BM is\ challenging in many cases, mainly due to the small thickness of this layer, the high reflectivity of the RPE that shadows parts of the BM, and the noise being present in the scans, which is often indistinguishable from the content of drusen and PEDs (Fig.~\ref{fig:amd_samples}). Due to these difficulties, currently many automated solutions either do not provide a segmentation of the BM or its segmentation is often inaccurate in retinal OCT with AMD, leaving this clinically relevant segmentation task unaddressed or under-explored. 

The state-of-the-art segmentation algorithms are based on Deep Learning (DL), which depends on a large amount of annotated training data that is often difficult and expensive to obtain. In addition, these models, when based only on textural features of the OCT images, may fail where the images contain artifacts due to the limitations in the scanning process, e.g., shadowing, eye-movement or low-resolution acquisition \cite{2018_Oktaya}. The introduction of prior knowledge about the target domain imposes constraints on the possible solutions, thus reducing the search space\cite{2016_Nosrati}.  Such knowledge can take several forms, including topology specification, distances between regions~\cite{2009_Garvin}, or shape models~\cite{2016_Nosrati}. Utilizing prior anatomical information for medical image segmentation has already been proven useful in order to obtain more accurate and plausible results, and with smaller training sets. They have been successfully applied among others to improve cardiac image segmentations~\cite{2018_Oktaya}, liver segmentation \cite{2019_Zheng_CONF} and retinal layer segmentation \cite{2006_Li, 2009_Garvin, 2021_He}.

Quantifying uncertainty of DL models is crucial for clinical applications in order to build trust in systems' prediction and at the same time for reducing the associated risks of downstream tasks relying on uncertain or incorrect results. This is particularly pertinent for image segmentation, where there is an inherent ambiguity in the reference, due to the limitations in the image acquisition processes and the subjectivity and complexity of the annotation task, resulting in variations between the manual annotations. However, DL segmentation methods tend to provide unrealistic overconfident predictions on these complex tasks, especially when they are applied on a different patient cohort or pathologies not observed during the training.

Having the above considerations in mind, in this paper, we propose a new deep learning method for automated segmentation of the BM. By using a probability distribution function to infer its spatial coordinates, together with a loss term incorporating anatomical priors which promotes smooth predictions, the method accounts for local morphological changes resulting from pathologies or acquisition artifacts, and is capable of identifying regions of potential segmentation failure. The acquired local uncertainty information is utilized in a post-processing step to further improve the segmentation in the areas of potentially erroneous segmentations. Large-scale multi-dataset experiments show the robustness of the developed model, which furthermore achieves the state-of-the-art performance on an external public dataset.

\subsection{Related works}
The goal of layer segmentation is to obtain anatomically coherent, smooth, and continuous retinal layer boundary surfaces. The first widely-used approach was to extract image features from the B-scans, which are then used by graph-based methods to estimate the surface positions. For instance, the IOWA Reference Algorithms \cite{2006_Li, 2014_Zhang} represented the OCT as a graph and the surface positioning was solved with dynamic programming algorithms, while guaranteeing the correct topological ordering, satisfying prior layer thickness constraints, and smoothing the results. The graph-based methods were later further improved in several works \cite{2010_Chiu,2013_Dufour,2014_Srinivasan}. Rathke\etal\cite{2017_Rathke_CONF} proposed a method using a probabilistic graphical model, which incorporated anatomical shape priors for OCT segmentation, including a post-processing fix for the BM. A drawback of these methods is that they rely on hand-crafted image features as the backbone for the graph construction and may perform poor in the presence of noise or other imaging artifacts, as well as severe pathologies. Several approaches attempted to improve on this by incorporating machine learning-based methods to estimate the cost function for the nodes of the graph \cite{2017_Fang, 2018_Kugelman,2019_Chen,2019_Santos_DISSERTATION}. For these types of approaches, the performance of the graph-search method is still tied to the quality of the initial probability map, and subject to predefined hard morphological constraints on layer thickness and smoothness variability.

With the advent of deep learning, U-Net \cite{2015_Ronneberger_CONF} and its variants became a dominant approach for medical image segmentation, including retinal layer segmentation. In particular, ReLayNet proposed in Roy\etal \cite{2017_Roy} presented a network architecture similar to U-Net, which represented nine retinal layers and possible fluid-filled pockets as distinct classes and predicted their pixel-wise locations. A deficiency of this algorithm is that it is not guaranteed to predict a single unique BM position in an A-scan. Sousa\etal \cite{2021_Sousa} uses a U-Net to create an initial segmentation followed by a CNN based edge detection network to further refine the results, while predicting one single location per A-scan.

A major weakness of these two deep learning methods is that they do not account for the  natural ordering of the layers, and consequently do not guarantee anatomically plausible results. The framework presented by He\etal~ \cite{2019_He_CONF} improves on this by predicting the surface positions using column-wise soft-argmax, thus ensuring that only one position is inferred per A-scan. Also, proper layer ordering is guaranteed with a topological module. They further improved their method in \cite{2021_He} by removing the fully-connected layers and hence requiring fewer parameters than in their previous work, while also evaluating the model performance on a BM segmentation task. Besides showing improvement against the state of the art they also showed that the surface connectivity is well constrained. However, they did not include  uncertainty estimation in their work and the method was not validated on AMD patients.

An alternative, not machine learning based approach was presented in Lou\etal \cite{2020_Lou}, which uses a mathematical model of the potential fluid energy in fluid mechanics. This method inherently guarantees the correct topological ordering and the smoothness of the predicted layers, however its performance is significantly lower than of the CNN based method, possibly because of the hard requirement of the algorithm, where the gray values on both side of the boundaries must be different.

Other researchers have focused on including uncertainty estimations coupled with the layer segmentation. The approach proposed by Orlando\etal~\cite{2019_Orlando_CONF} predicts the photoreceptor layer, and they perform Bayesian inference through Monte Carlo sampling using the dropout in a modified U-Net architecture. They investigated the correlation between the measured uncertainty and the segmentation performance, although only on a B-scan and volume level.
In addition to using the same approach to quantify uncertainty, the framework proposed by Sedai\etal \cite{2018_Sedai_CONF} uses a fully convolutional network that learns to output the aleatoric uncertainty which it was observing. The network performed comparably to the state of the art, but the relation between the uncertainty and the segmentation displacement, essential for the clinical applicability, was not covered in the work. 

A problem common to all the aforementioned methods is that they lack one or more critical components for successful clinical translation in AMD. \NV{Either they are not able to robustly predict the BM, or they were not validated on all stages of the AMD, with different acquisition settings, or they do not provide an uncertainty estimation required for achieving reliable, trustworthy segmentation methods. }

\subsection{Summary of contributions}
In this paper, we propose a novel deep learning method for segmentation of the BM layer from retinal OCT scans of patients with AMD. The main contributions of this work are the following:

\begin{itemize}
    \item A new curvature loss term to encode a shape anatomic prior of the BM. This improves the model's robustness and the ability to detect the BM in low-contrast areas, resulting in anatomically more plausible solutions.
    \item Uncertainty quantification on A-scan, B-scan and OCT volume level to detect possible mis-segmentations, which can then be automatically corrected in a post-processing step.
    \item \NV{Large-scale evaluation of the proposed method, across all three AMD stages and on images acquired with different OCT devices from various patient cohorts reflecting a real-world clinical setting, as well as on an external public test set, proving the strong generalization capability of the solution.}
\end{itemize}

\section{Methods}
\label{sec:methods}

\begin{figure*}[tb]%
    \centering
    \includegraphics[width=\textwidth]{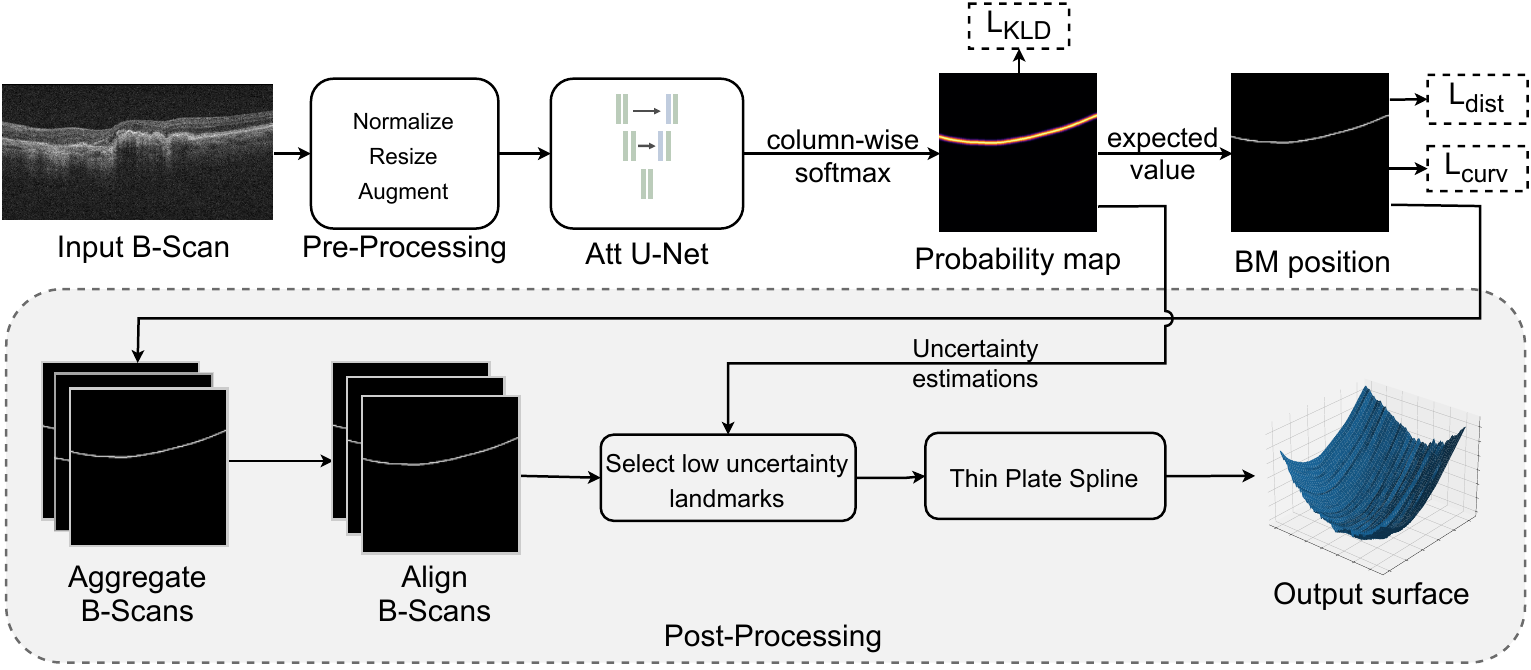} %
    \caption{An overview of the proposed method. The input is a single B-scan, which is pre-processed. The Att. U-Net predicts a probability map of equal size as the input. Each column contains a probability distribution of the possible BM location. The actual BM position is calculated from the expected value. Besides matching the correct position, a curvature constraint is imposed to learn the expected shape of the layer. During inference, the calculated probability maps from a volumetric scan are aggregated, aligned, and using random control points with low uncertainty a Thin Plate Spline surface is calculated to replace the highly uncertain values.}%
    \label{fig:graphical_abstract}%
\end{figure*}

The proposed deep learning method is designed to provide an anatomically coherent segmentation of the BM. The input to the network is a pre-processed B-scan and the single channel output contains column-wise (A-scan-wise) a probability distribution of the BM position. The BM positions are then regressed from the expected values of the distributions. During the training, curvature constraints are imposed to provide an inductive bias on the correct shape of the surface even in the areas, where it is hardly visible. During inference, volumetric predictions are obtained by combining the B-scan-wise segmentations and afterwards replacing uncertain regions with interpolated values.

\subsection{Regressing BM position with anatomy-aware probability distributions}
All of the B-scans were resized regardless of the initial resolution to $512 \times 512$ pixels using bilinear interpolation. The backbone image segmentation network is an Attention U-Net \cite{2018_Oktay} with five downsampling and upsampling layers with 32, 64, 128, 256 and 512 channels, respectively. The input is a $512\times 512 \times 1$ image and and the output layer has a single channel of the same size as the input. Every convolutional layer is followed by a dropout layer with 0.2 drop probability. We used LeakyReLU as activation function in the hidden layers. The output of the network is a column-wise probability map, in which each column represents a probability distribution, where the value of each row corresponds to the probability of the BM at this position. The target distribution is modeled to be a Gaussian distribution to better accommodate for possible pixel-wise imprecisions of the annotations. This allows the network to identify several adjacent positions within a columns as possible candidates.

We used a loss function consisting of three terms, each focusing on a different aspect of the segmentation task: (i) a term for regressing correct position of the BM, (ii) a term providing a pixel-wise supervision per image columns to weakly enforce a Gaussian distribution of the probability mass function, used later for the uncertainty estimation, and (iii) a curvature term which introduces the anatomical shape prior in the training itself, and regularizes the curvature of the predicted BM as well as enforcing its continuity. 

\subsubsection{Surface position regression with a probability mass function}
\label{sec:layer_regression}

To regress the coordinates of the BM, the column-wise expected value of the probability map is calculated, similarly to the models proposed by He\etal\cite{2019_He_CONF, 2021_He}. 

Let $Y$ be a random variable corresponding to the $y$-coordinate (position) of the BM, and $X$ is the position of an A-scan. We aim to assess the expected BM position given the A-scan $x$:

\begin{equation}
\hat\mu_{Y\mid x} = \sum_y{y\cdot P(Y=y\mid X = x)}
\label{eq:expected_value}
\end{equation}

The probability mass function (PMF) $P(Y\mid X)$ is estimated with the neural network. To ensure that $P(Y\mid X)$ is a proper PMF, we perform a column-wise softmax activation over the network outputs.

The PMF, thus BM position, is learned with the help of mean squared error (MSE) loss between the predicted BM location $\boldsymbol{\hat\mu}$ and the reference standard $\boldsymbol{\mu}$:

\begin{equation}
    \mathcal{L}_1 = \sum_{x}P(x)\left(\boldsymbol{\hat\mu_{Y\mid x}} - \boldsymbol{\mu_{Y\mid x}}\right)^2
\end{equation}

\noindent We assume that all A-scans are equally important and $X\sim~U[1,N]$, where $N$ is a number of A-scans in a single B-scan.

\subsubsection{Regularization of the distribution}
As proposed in Nibali\etal \cite{2019_Nibali_CONF}, we introduce a regularization term to guide the model to match the output distribution to a target Gaussian probability mass function for each column and thus introducing a pixel-level supervision. We opted to use Kullback-Leibler divergence ($D_{\text{KL}}$) \cite{1951_Kullback} instead of Jensen-Shannon divergence as used in \cite{2019_Nibali_CONF} as it resulted in similar performance with lower computational cost.

Thus, we introduced the second loss term:

\begin{equation}
\begin{aligned}
    \mathcal{L}_2 &= D_{\text{KL}}\left(\mathbf{ P(Y\mid X)} \parallel    \mathbf{T (Y\mid X)} \right) \\ &=\sum_x\sum_{y}{\mathbf{P}\left(y, x\right) \ln \frac{\mathbf{P}\left(y\mid x\right)}{\mathbf{T}\left(y\mid x\right)}} ,
    \label{eq:kl_divergence}
\end{aligned}
\end{equation}

where $\mathbf{ P}$ is the probability map from the network output and $\mathbf{T}(Y\mid x)\sim N(\mu_{Y\mid x},\sigma)$ is the target probability map, containing a Gaussian distribution for each column $x$ with the mean $\mu_{Y\mid x}$ being the reference standard location, and having standard deviation $\sigma$.
This loss term penalizes output distributions different from the target Gaussian distribution.
\if false
Fig.~\ref{fig:with_without_kld}. illustrates the effect of using KL-divergence to match the output to a column-wise Gaussian distribution. 

\begin{figure}[tb]
    \begin{minipage}{.49\textwidth}
    \centering
        \includegraphics[scale=0.35]{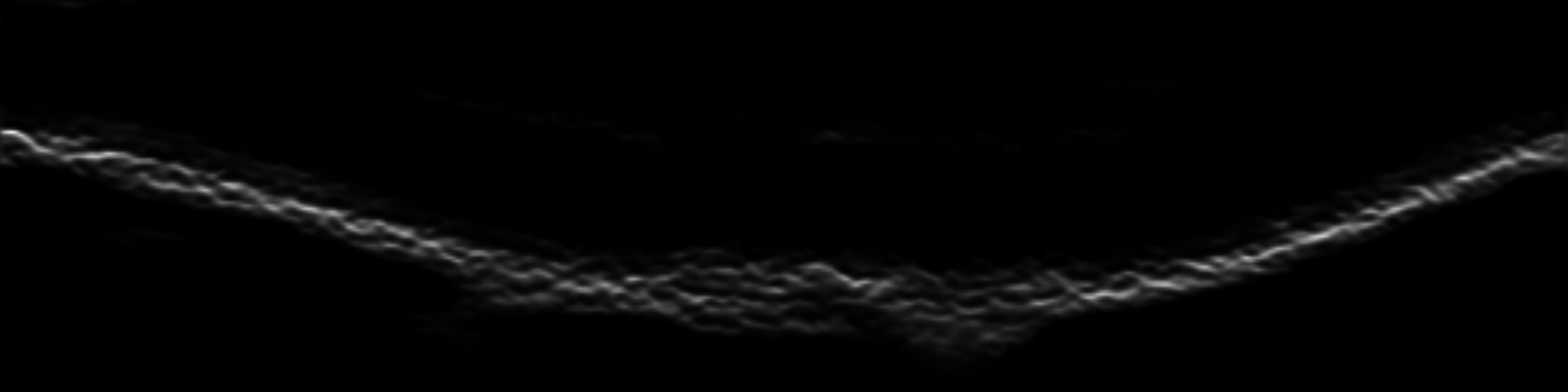}
        \caption*{(a) without distribution regularization}
    \end{minipage}
    \hfill
    \begin{minipage}{.49\textwidth}
        \centering
        \includegraphics[scale=0.35]{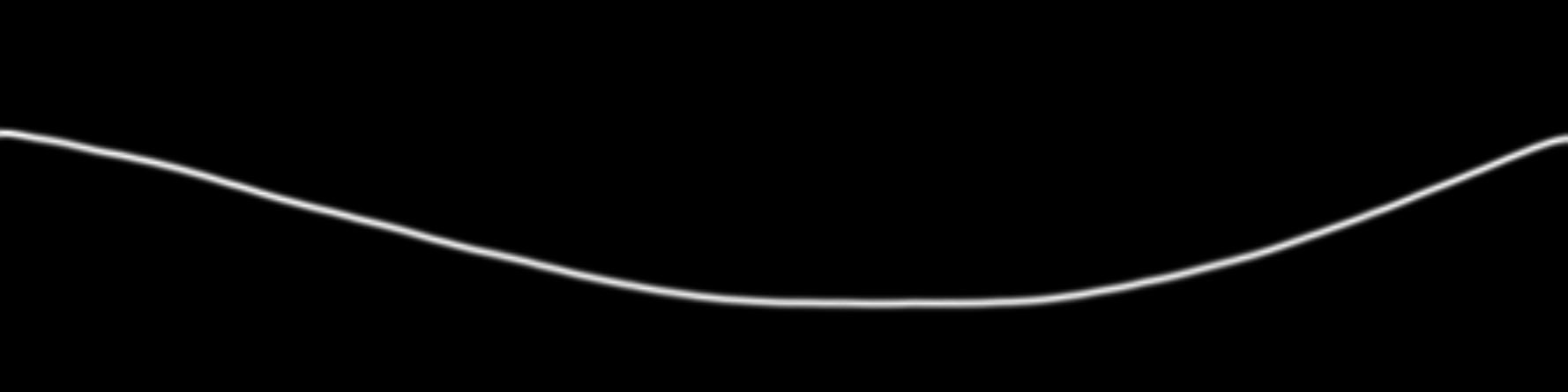}
        \caption*{(b) with distribution regularization}
    \end{minipage}
    \caption{Sample outputs generated by a network trained using (a) $\mathcal{L}_1$ only, and (b) using $\mathcal{L}_1 + \mathcal{L}_2$. While the expected value is the same for both of the outputs, (b) resembles more closely a Gaussian distribution per A-scan.}
    \label{fig:with_without_kld}
\end{figure}
\fi

\subsubsection{Curvature preserving loss function}

We further improve the quality of the segmentation by incorporating anatomical prior on the smoothness of BM with an extra term in the loss function. This allows reducing segmentation errors created by pathologies and scanning artifacts that reduce the visibility of the BM. Specifically, we introduce a novel term that leads the network to match the slope of the predicted surface to that of the manually annotated reference. Additionally, while the surface position regression and $D_{KL}$ is applied column-wise, this term introduces an explicit connection between the neighboring columns.

The absolute curvature $\kappa$ of a function $f(x)$ is defined as follows \cite{1998_Morris_BOOK}:

\begin{equation}
\kappa(x) = \frac{|f''(x)|}{\left(1+{f'(x)}^2\right)^\frac32}
\end{equation}

\noindent Since we predict the y-coordinate of the BM only at discrete A-scan positions, where $f(x) = \mu_{Y\mid x}$, we approximate the first and the second derivatives with finite differences:

\begin{equation}
    f'(x) \approx f\left(x+\left\lfloor\tfrac12h\right\rfloor\right)-f\left(x-\left\lfloor\tfrac12h\right\rfloor\right)
    \label{eq:first_deriv}
\end{equation}
\begin{equation}
    f''(x) \approx -f\left(x+\left\lfloor\tfrac12h\right\rfloor\right) + 2f\left(x\right)-f\left(x-\left\lfloor\tfrac12h\right\rfloor\right),
    \label{eq:second_deriv}
\end{equation}

\noindent where $h$ is the window-width and $x$ is the index of the $x$th A-scan.
The first and the second derivative can be calculated using a 1D-Convolution with the kernels of length $h$: $\rvect{1 & \cdots & -1 }$ and $\rvect{-1 & \cdots & 2 & \cdots  & -1}$, respectively, with the omitted values being 0.

We define the loss as the mean absolute error (MAE) between the point-wise curvature in the manual reference $\boldsymbol{\mu}$ and the predicted position $\boldsymbol{\hat\mu}$:

\begin{equation}
    \mathcal{L}_3 = MAE\Big(\kappa_h\left(\boldsymbol{\mu}\right), \kappa_h\left(\boldsymbol{\hat\mu}\right)\Big)
\end{equation}

The  window size $h$ has a considerable effect on the weight of $\mathcal{L}_3$ term in the complete loss.
A very large $h$ allows to capture low and high curvatures on curved sections of different lengths while being robust to noise. However, the estimation of the curvature is only possible on columns at least $\frac12h$ away from the left/right image borders, and thus large windows may reduce the influence of $\mathcal{L}_3$ during training. On the other hand, a small $h$ only allows to penalize short sections with high curvature, failing to capture longer low curvature curves.

The complete loss term is thus:
\begin{equation}
    \mathcal{L} = \alpha\mathcal{L}_1 + \beta\mathcal{L}_2 + \gamma\mathcal{L}_3
    \label{eq:weighting}
\end{equation}
\NV{where $alpha$, $\beta$ and $\gamma$ are weighting hyper-parameters.}

\subsection{Uncertainty estimation}
Due to the deficiencies in the image acquisition process and the complex anatomical deformations caused by retinal diseases, the segmentation method is not always able to correctly identify the BM position. Thus, it is important to have the ability to quantify the uncertainty observed during inference time. As the output of the network is a probability map, if the network is certain at the location of the BM, it assigns higher probabilities only to the positions lying very close to the expected value. In contrast, if it is uncertain, it assigns lower probabilities over many positions lying further apart from each other.

In order to quantity this uncertainty, we calculate the standard deviation $\hat\sigma_{Y\mid x}$ for every A-scan $x$, analog to the calculation of the expected value in \eqref{eq:expected_value}:

\begin{equation}
\hat\sigma_{Y\mid x} = \sqrt{\sum_{y}{\left(y - \hat\mu_{Y\mid x}\right)^2 \cdot P\left(y\mid x\right)}}
\end{equation}

\subsection{Post-processing with thin plate splines}

We use a thin plate spline (TPS) \cite{1977_Duchon_CONF} interpolation on the volumetric 3D OCT scan to estimate a viable position of the BM in the uncertain areas. The TPS finds an interpolating surface with a set of control points to assure minimum bending of the BM. With TPS, we replace the uncertain BM positions with their interpolated values.

\paragraph{\NV{B-scan alignment}}TPS interpolation across B-scans assumes a topographically smooth surface under the condition that the neighboring B-scans are aligned along the axial direction, which is not always assured due to eye motion artifacts. While Spectralis devices contain a built-in alignment method, Cirrus scans are prone to this effect. \NV{\st{With this in mind} To align the B-scans prior to the TPS interpolation}, we calculate a displacement vector $\delta_j$ for each B-scan $j$ in a scan, that corresponds to the difference between the mean position of the BM in $j$ and a prespecified reference position $R$.

\begin{equation}
    \delta_j = \frac{1}{N} \sum_{x = 1}^N{\hat\mu_{Y|x,j}} - R,
    \label{eq:alignment}
\end{equation}
where $N$ is the number of A-scans in a B-scan $j$. We used $R=\frac12H$, where $H$ is the height of the B-scans. 

\paragraph{\NV{TPS interpolation}} The TPS is then calculated on the centered BM layer after which the BM is realigned to the original average positions:
\begin{equation}
    \boldsymbol{\hat\mu'} = \text{TPS}\left(\boldsymbol{\tilde\mu}  - \boldsymbol{\delta}\right) + \boldsymbol{\delta}.
\end{equation}
The number of control points and the selection of the rigidity parameter of TPS 
have a considerable effect on the smoothness and the resulting accuracy of the interpolated surface. Using more control points results in a larger computational cost, and makes the result more prone to erroneous control point positions and B-scan misalignment artifacts. A larger rigidity allows the interpolation to deviate more from the control points. The rigidity parameter and the number of control points were determined via cross-validation on the validation set and were defined to 0.05 and 1024, respectively.

\begin{figure}[tb]
    \begin{minipage}{.24\textwidth}
    \centering
        \includegraphics[width=0.8\textwidth]{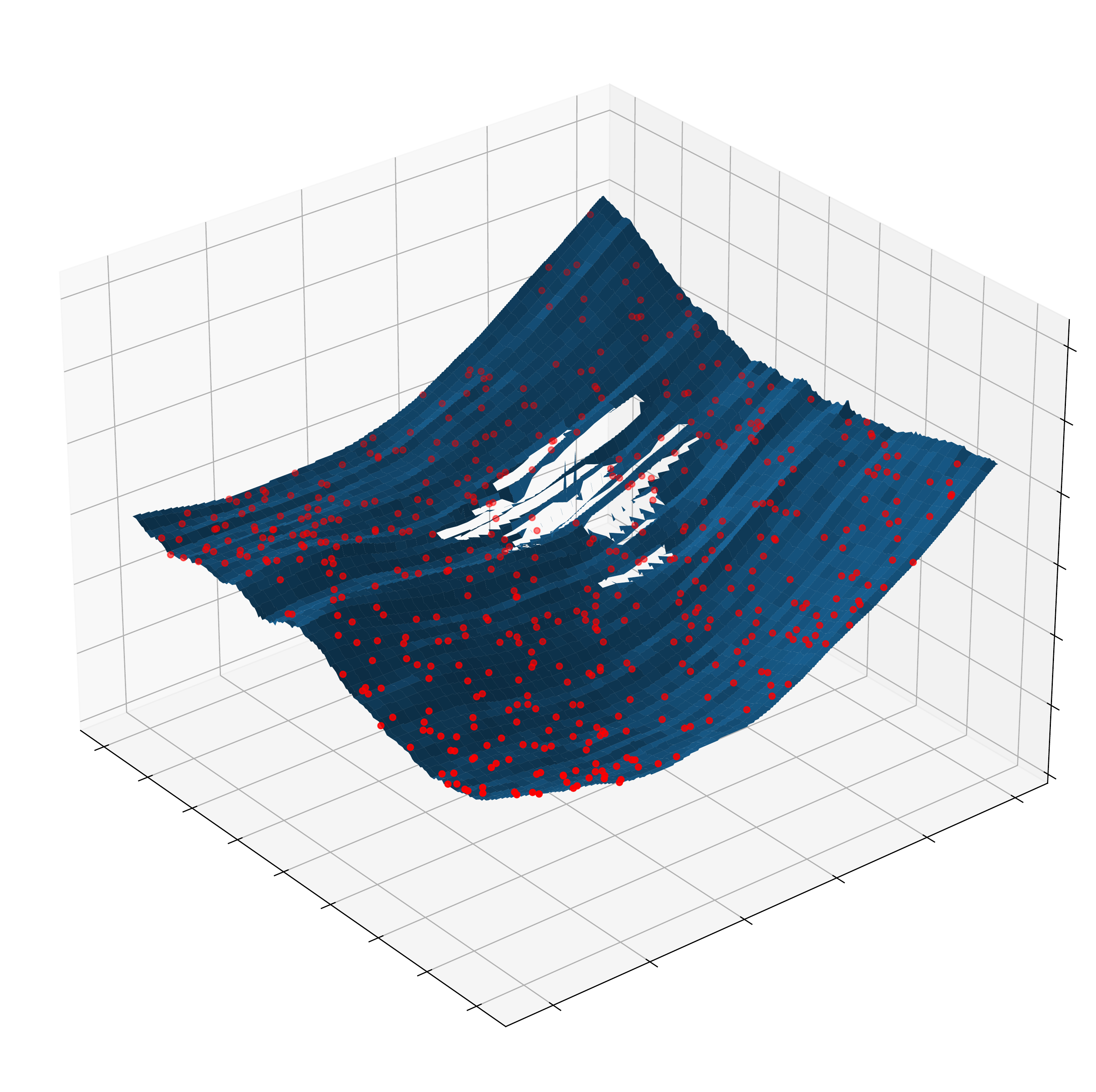}
        \caption*{(a) Before TPS}
    \end{minipage}
    \hfill
    \begin{minipage}{.24\textwidth}
        \centering
        \includegraphics[width=0.8\textwidth]{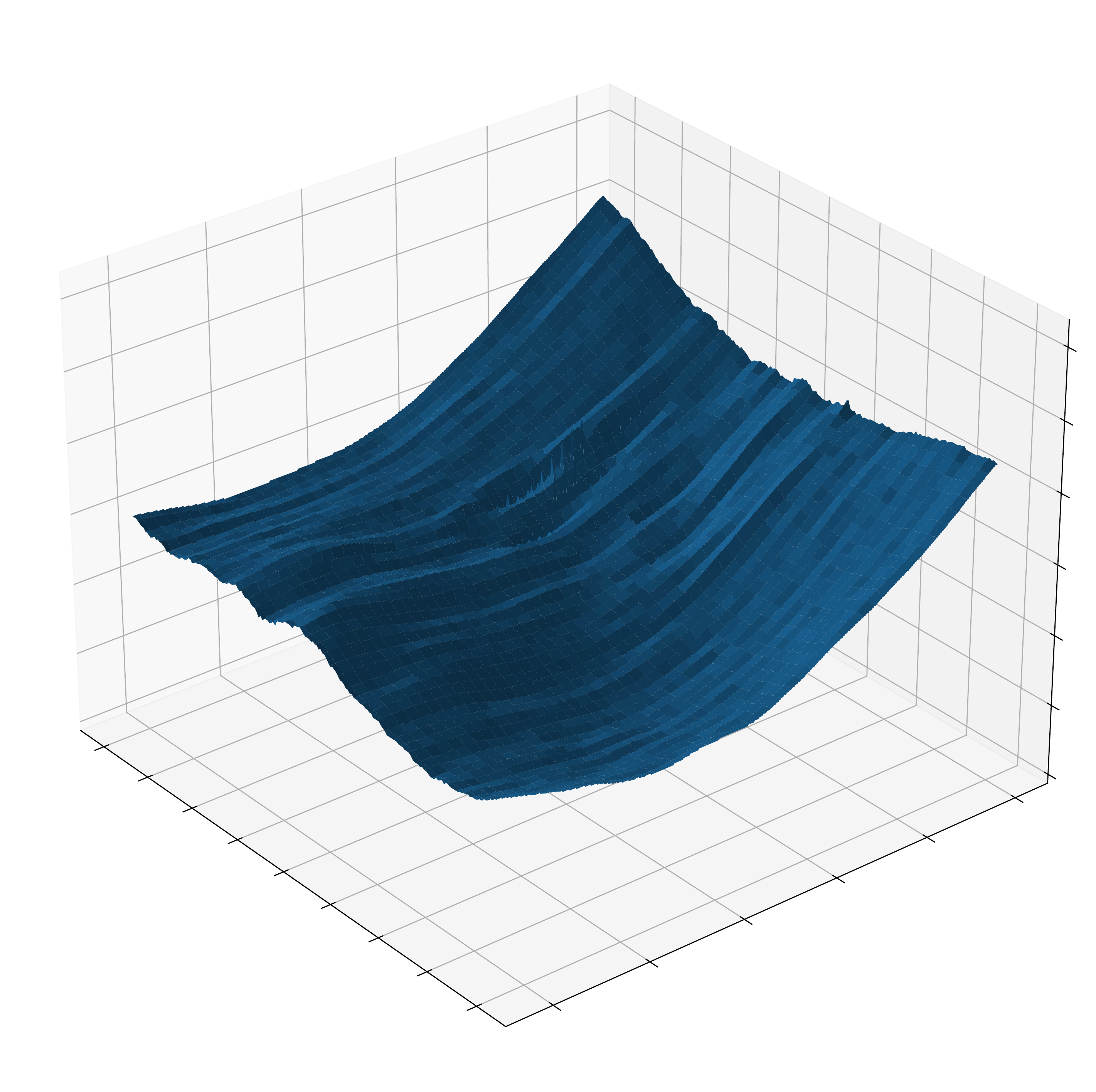}
        \caption*{(b) After TPS}
    \end{minipage}
    \caption{The left image shows the surface consisting of low uncertainty predictions, and the randomly selected control points for the TPS with red. The surface containing the interpolated parts is depicted on the right.}
\end{figure}

\section{Experimental setup}
\label{sec:experiments}

In this section we give an overview of the datasets we used to train, validate and test our method. We also describe the pre-processing and data augmentation steps, the training setup, and the evaluation metrics considered. Finally, a description of the performed experiments is provided.

\subsection{Datasets}

The proposed method was trained and evaluated on an internal dataset consisting of 1,449 volumetric OCT scans of 478 eyes from 386 patients, covering all three stages of AMD: iAMD, nAMD and GA. The analysis adhered to the tenets of the Declaration of Helsinki, and the approval was obtained by the
Ethics Committee of the Medical University of Vienna (Nr 1246/2016). 71\% of the scans were acquired with Spectralis devices (Heidelberg Engineering, Heidelberg, Germany) consisting of $(512-1024) \times (23-97) \times 496$ voxels with a size of $ (5.74-11.46) \times (60.90-232.48) \times 3.87~\upmu\text{m}^3$ comprising a volume of $(5.6-8.6) \times (4.1-7.4) \times 1.9~\text{mm}^3$. The remaining 29\% was acquired with Cirrus devices (Carl Zeiss Meditec, Dublin, CA, USA) consisting of $(200-512) \times (128-200) \times 1024$ voxels with a size of $ (11.7-30.1) \times (30.1-47.2) \times 1.95~\upmu\text{m}^3$ comprising a volume of $6 \times 6 \times 2.0~ \text{mm}^3$. The BM reference standard locations were created by the IOWA Reference Algorithm \cite{2006_Li, 2014_Zhang} which were manually corrected by retinal experts afterwards. The dataset was split up on a patient-level in 80\% for training, and 10\%-10\% for validation and test. \NV{In order to keep the size of the training set manageable, 10  B-scans were initially randomly selected from each OCT volume and this reduced training set was used for our experiments.} The validation and test sets contained all of the B-scans available. Table~\ref{tab:testset_splitup} gives an overview of the dataset and the split across the three sets.

\begin{table}[tb]
    \centering
    \begin{tabular}{|l||c|c|c|}
    \hline
         \diagbox[width=2cm]{\textbf{Disease}}{\textbf{Vendor}}             & \textbf{Cirrus} & \textbf{Spectralis} & {\textbf{Total}} \\ \hhline{|=||=|=|=|}
{\textbf{iAMD}}  & N/A             & 507 / 63 / 65                  & {\textbf{507 / 63 / 65}}            \\ \hline
{\textbf{nAMD}}  & 158 / 19 / 19              & 82 / 11 / 10                  & {\textbf{240 / 30 / 29}}            \\ \hline
{\textbf{GA}}    & 173 / 21 / 25              & 248 / 28 / 20                  & {\textbf{421 / 49 / 45}}            \\ \hline
{\textbf{Total}} & \textbf{331 / 40 / 44}              & \textbf{837 / 102 / 94}                  &{\textbf{1168 / 142 / 139}}              \\ \hline
\end{tabular}%
    \caption{Number of volumetric scans (train/validation/test) per vendor and disease of the internal dataset.}
    \label{tab:testset_splitup}
\end{table}

\subsection{Generalization to an external test test and domain shift}
To additionally evaluate the generalizability of our trained network, we carried out an evaluation on a public external test set, allowing to also compare the results with the reported state of the art. For this purpose, we used the openly available Duke SD-OCT dataset \cite{2014_Farsiu}, the world's largest online annotated SD-OCT dataset, containing 38,400 annotated B-scans of 259 iAMD patients and 115 healthy subjects, coming from four external clinical centers of the Age-Related Eye Disease Study 2 (AREDS2) Ancillary SD-OCT Study.
The scans were acquired with a Bioptigen SD-OCT device (Research Triangle Park, NC) consisting of $1000 \times 100 \times 512$ voxels with a size of $6.7 \times 67.0 \times 3.24$~$\upmu\text{m}^3$ comprising a volume of $6.7 \times 6.7 \times 1.7$~$\text{mm}^3$. Our method was never trained on these scans, allowing to assess the model's generalization under an image domain shift.

\subsection{Experimental settings}
\subsubsection{Pre-processing}
The input volumes were normalized B-scan-wise to zero mean and unit standard deviation. Due to the movement of the patient and other scanning artifacts, not all the A-scans have a BM defined. If in a B-scan the BM was undefined in more than 20\% of the A-scans, the B-scan was removed from the training set. For the validation and test sets, B-scans that contained no definition of the BM were not considered.

\subsubsection{Implementation and training details}
The B-scans during the training were augmented with random horizontal flipping, rotated by a random angle between $-20^{\circ}$ and $20^{\circ}$,  distorted along a random sinusoidal wave to represent scanning artifacts, and randomly shifted vertically. Each augmentation was applied with a probability of 0.3.

\NV{In preliminary experiments on the validation set, we tested different window sizes $h$ (Eq.~\ref{eq:first_deriv} and \ref{eq:second_deriv}) We tested the values $3, 5, 7, 11, 21, 31, 51$, and found that $h = 21$, approximately $4\%$ of the image width, leads to the best performance among these values. Likewise, we tested different $\sigma$ (Eq.~\ref{eq:kl_divergence}) values of the target Gaussian distribution between 0.5 and 2.0, and found no significant difference in the performance, thus using $\sigma = 1.0$ for the sake of simplicity. In our experiments, similarly to \cite{2019_He_CONF, 2021_He}, we set $\alpha = \beta = 1.0$ in Eq.~\ref{eq:weighting}. For $\gamma$ (Eq.~\ref{eq:weighting}) we tried the values $0.5, 1, 2, 5, 10, 20$, and we found no statistically significant difference in the performance of the fully trained models, but higher values have led to faster initial convergence. In our main experiments we used $\gamma = 1.0$.}

Our method and the baseline were trained with a batch size of 10 using the Adam optimizer \cite{2014_Kingma}, with an initial learning rate of 0.05, which was then lowered in the 3\ts{rd}, 7\ts{th}, 10\ts{th}, 30\ts{th} and then every further 20 epochs by 30\%. We trained the network for 200 epochs and we selected the model with the lowest RMSE on the validation set. The dropout probability was 0.2.

The training was performed in a mixed-precision setting on an Nvidia GeForce RTX 2080 Ti GPU, with an Intel Xeon Silver 4114 CPU with 16 cores, under CentOS 7, using Python 3.8.8 and PyTorch 1.8.0. On such a setup, the training lasted approximately 32 hours. 
\NV{The testing was done on the same setup as the training. The inference of the B-scans and the TPS post-processing lasts about 6 seconds from the input of the OCT volume to the output of the predicted 3D surface.}

\subsubsection{Evaluation metrics}

The proposed method is evaluated in terms of the A-scan-wise distance between the automated segmentation and the manual reference using the common metrics, mean absolute error (MAE), and the root mean squared error (RMSE) that penalizes more strongly the large segmentation errors. Furthermore, there is also particular interest in examining the smoothness and continuity of the resulting segmentation. Discontinuities of the BM may result from wrong segmentations due to some pathologies or bad quality scans and they represent anatomically implausible results and could bias downstream measured clinically relevant parameters. To  evaluate the smoothness of the segmented BM, we plotted a histogram,
containing the distance of adjacent columns $\mu(x + 1) - \mu(x)$, where $\mu(x)$ is the position of the BM at an A-scan $x$. 

\subsection{Experiments}
\subsubsection{Comparison with baseline and ablation studies}
\label{sec:experiments_baseline}

\NV{We compared the performance of our model with adaptions of state-of-the-art retinal OCT layer segmentation methods,  DexiNed \cite{2021_Sousa}, ReLayNet \cite{2017_Roy} and Bayesian Fully Connected Dense Network (BFC-DN) \cite{2018_Sedai_CONF}. DexiNed is an edge detection network that produces pixel-wise edge maps. As a consequence, it does not directly predict a boundary coordinate per A-scan. To obtain the final coordinates we generated predictions using both column-wise argmax and the expected value as in Eq.~\ref{eq:expected_value}. We found that the latter solution performed better and we used those results to compare with our method.}

\NV{ReLayNet \cite{2017_Roy} is a pixel-wise retinal OCT layer segmentation network. The final layer consists of two channels for the two output classes corresponding to the areas below and above the BM. Softmax is used to estimate the probability of a pixel belonging to either of the two classes. We adapted ReLayNet by using Attention U-Net as opposed to standard U-net, which we empirically found to perform better for the task in study. At each pixel, the class with higher priority was selected. As a post-processing step, in each class only the largest connected component was kept to remove possible wrong classifications. The position of the BM is the border of the two classes.}

\NV{Bayesian Fully Connected Dense Network (BFC-DN) \cite{2018_Sedai_CONF} is also an U-net-like method that performs pixel-wise retinal OCT segmentation while accounting for the prediction uncertainty. Similarly to ReLayNet, the network is trained to segment the regions above and below the BM. We used the same post-processing step as with \cite{2017_Roy} to improve the results.}

In addition, an ablation study was conducted in order to assess the effectiveness of the curvature term in the loss, and the TPS in contrast to the basic version of our method, consisting of only the first two loss terms $\mathcal{L}_1$ and $\mathcal{L}_2$. This is similar to the regression model proposed in He\etal\cite{2021_He}.
We compared our proposed algorithm with the baseline methods in terms of MAE, RMSE and smoothness histogram.
We used the Wilcoxon signed-rank test to test whether the proposed method and its components represent a significant improvement.

\subsubsection{Performance across AMD stages and OCT devices}
To assess the viability of the method in a real-world clinical setting, model evaluation was performed for the ablation study models across the three major disease stages and scans acquired with different OCT devices.

\subsubsection{Validation on an external test set under domain shift}
\NV{To compare our method with the current state-of-the-art solutions and to further test the generalization ability of the network we evaluated the trained model on the external public dataset from Duke.} For comparison with the existing literature, results are reported using the MAE and the standard deviation.

\subsubsection{Uncertainty estimation}
To assess the correlation between the observed uncertainty and the displacements of the predicted BM positions from the reference standard, we report the Spearman correlation coefficient (SCC) on A-scan, B-scan and volume level. We also inspected the 2D histogram between the displacements per A-scan against the standard deviation per A-scan in order to evaluate how well the uncertainty measure can be used to predict the quality of the segmentation.

\section{Results}
\label{sec:results}

\subsection{Comparison with baseline and ablation studies}

\begin{figure*}[htb]
    \begin{minipage}{.24\textwidth}
    \centering
        \includegraphics[width=1.0\textwidth]{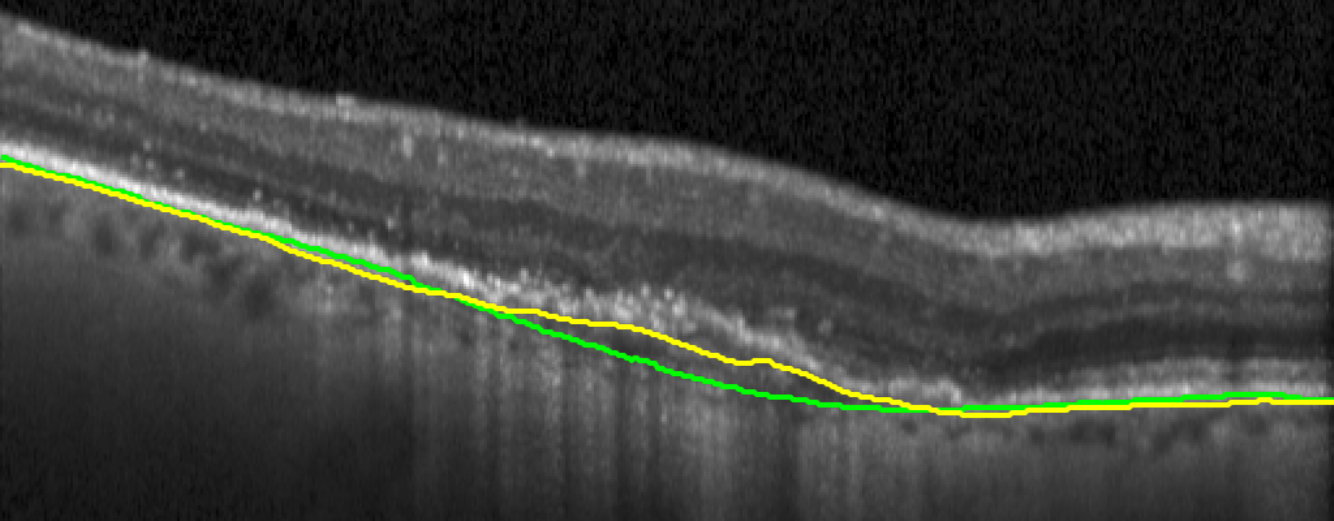}
    \vspace{0.01cm}
    \end{minipage}
    \begin{minipage}{.24\textwidth}
    \centering
        \includegraphics[width=1.0\textwidth]{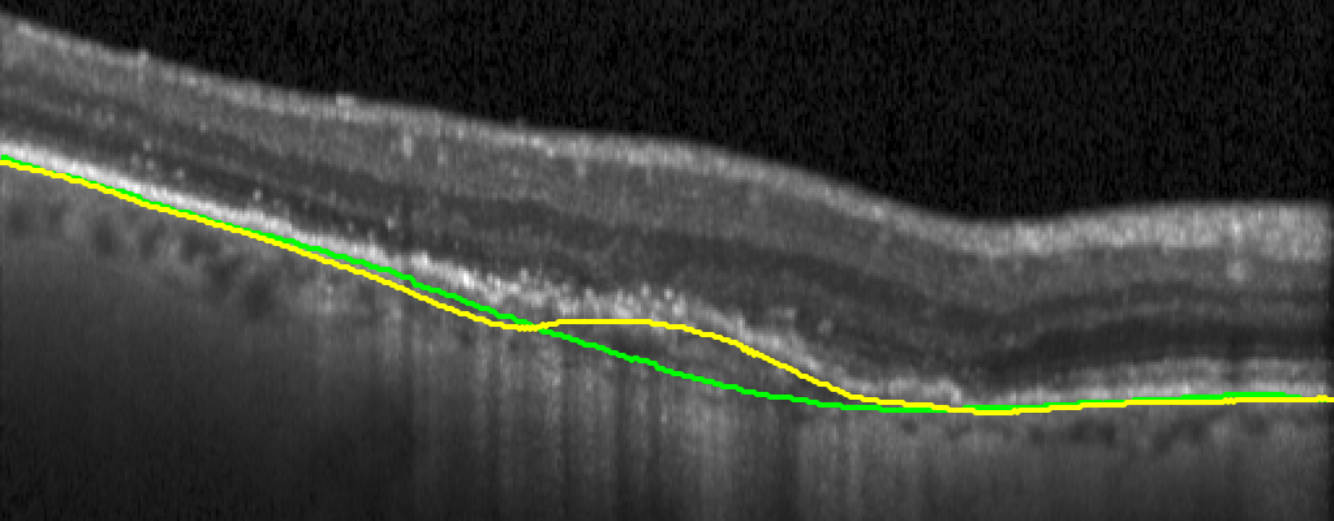}
    \vspace{0.01cm}
    \end{minipage}
    \hfill
    \begin{minipage}{.24\textwidth}
        \centering
        \includegraphics[width=1.0\textwidth]{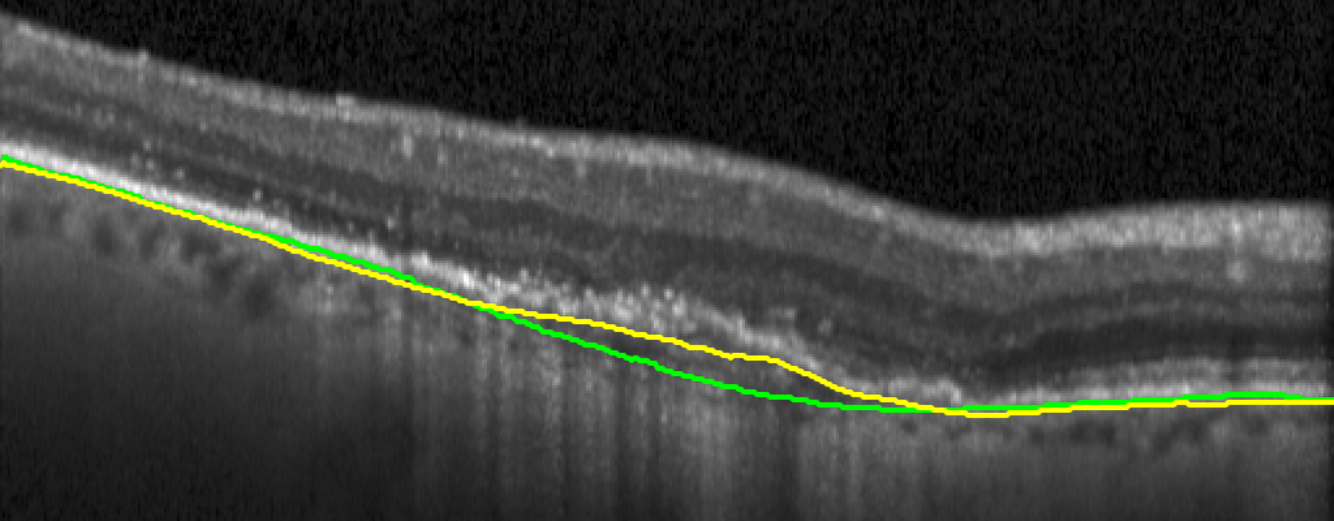}
    \vspace{0.01cm}
    \end{minipage}
    \hfill
    \begin{minipage}{.24\textwidth}
        \centering
        \includegraphics[width=1.0\textwidth]{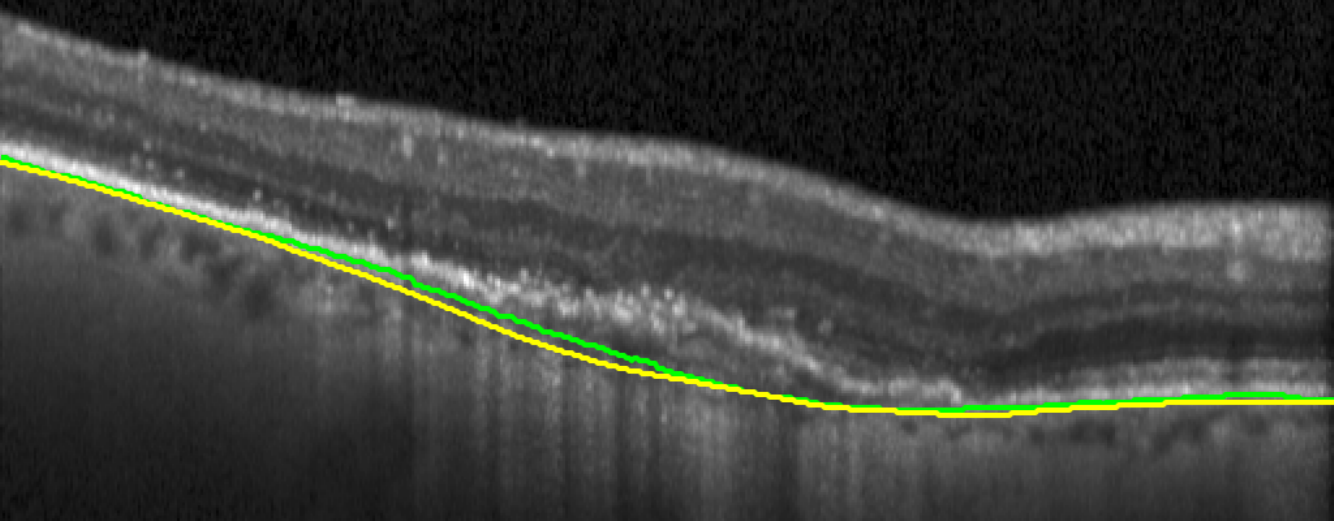}
    \vspace{0.01cm}
    \end{minipage}
    \begin{minipage}{.24\textwidth}
    \centering
        \includegraphics[width=1.0\textwidth]{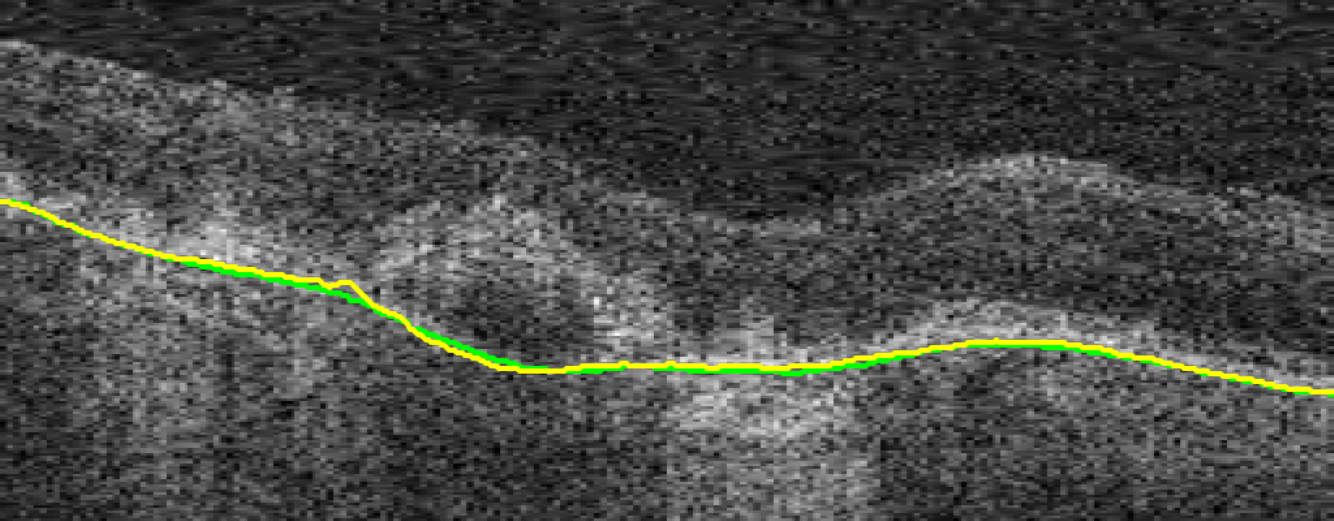}
    \vspace{0.01cm}
    \end{minipage}
    \hfill
    \begin{minipage}{.24\textwidth}
    \centering
        \includegraphics[width=1.0\textwidth]{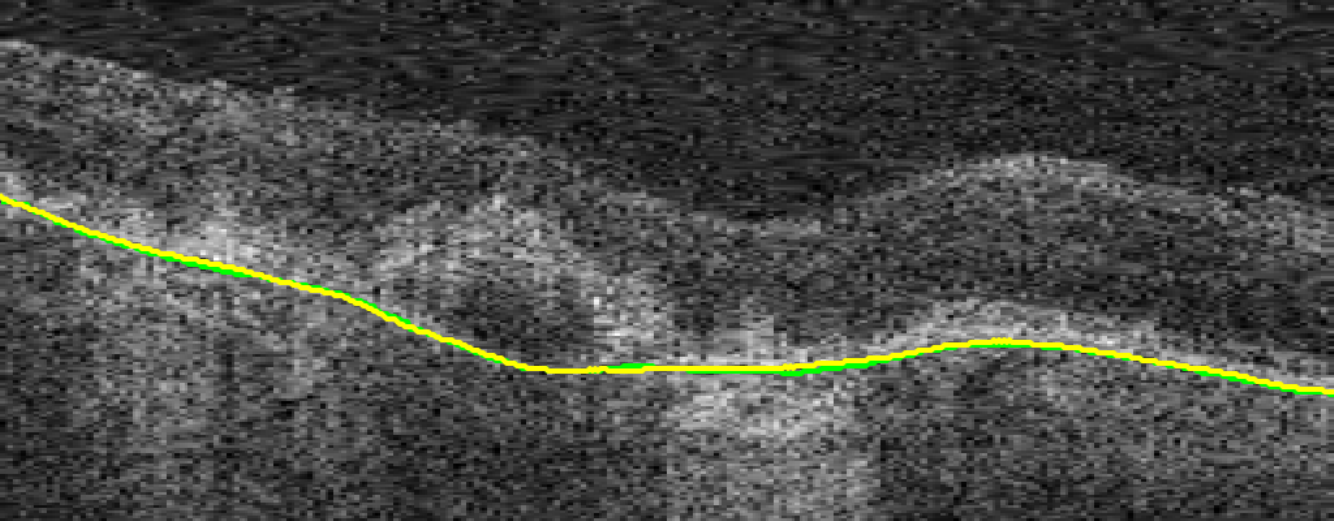}
    \vspace{0.01cm}
    \end{minipage}
    \hfill
    \begin{minipage}{.24\textwidth}
        \centering
        \includegraphics[width=1.0\textwidth]{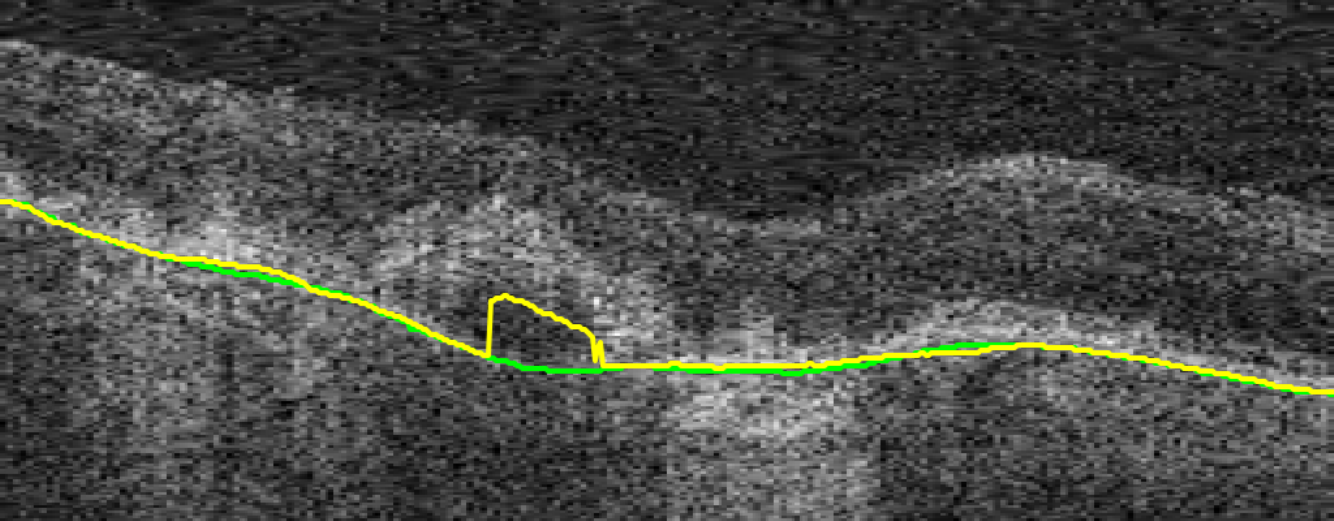}
    \vspace{0.01cm}
    \end{minipage}
    \hfill
    \begin{minipage}{.24\textwidth}
        \centering
        \includegraphics[width=1.0\textwidth]{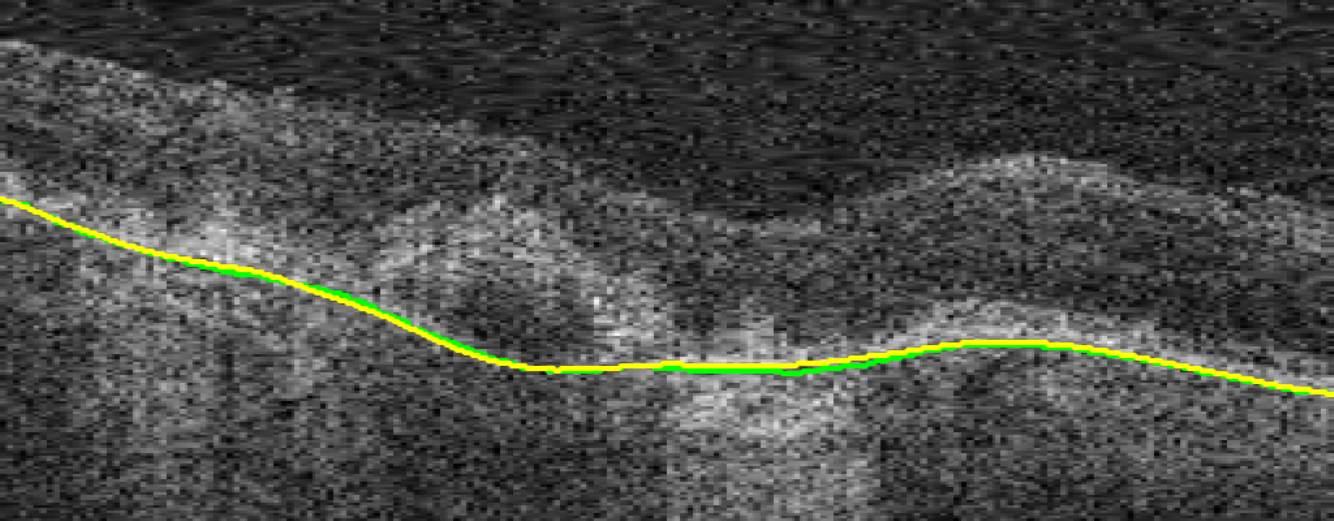}
    \vspace{0.01cm}
    \end{minipage}
    \begin{minipage}{.24\textwidth}
    \centering
        \includegraphics[width=1.0\textwidth]{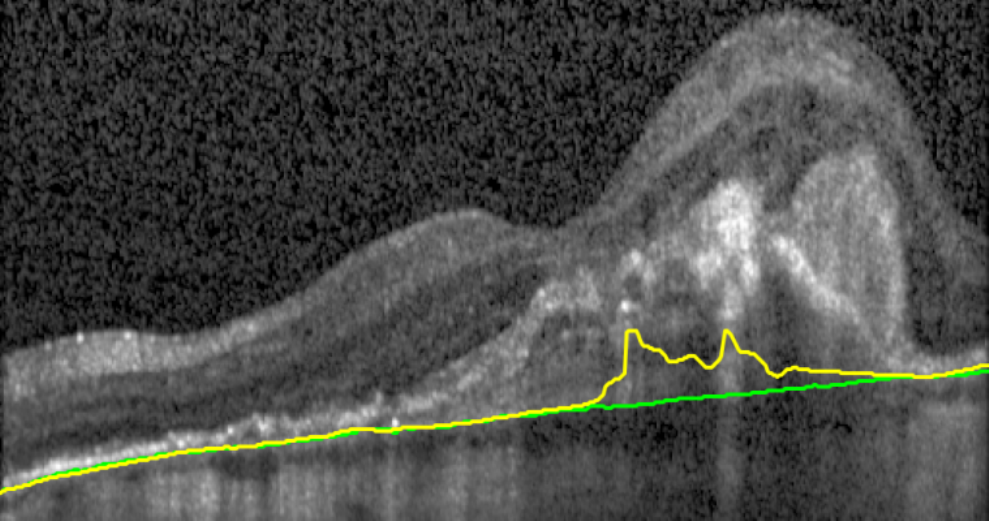}
    \vspace{0.01cm}
    \end{minipage}
    \hfill
    \begin{minipage}{.24\textwidth}
    \centering
        \includegraphics[width=1.0\textwidth]{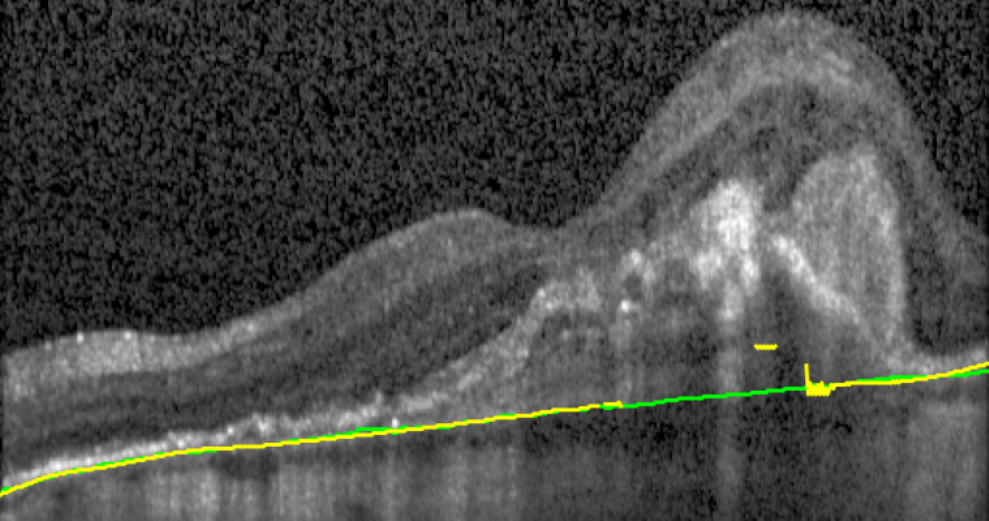}
    \vspace{0.01cm}
    \end{minipage}
    \hfill
    \begin{minipage}{.24\textwidth}
        \centering
        \includegraphics[width=1.0\textwidth]{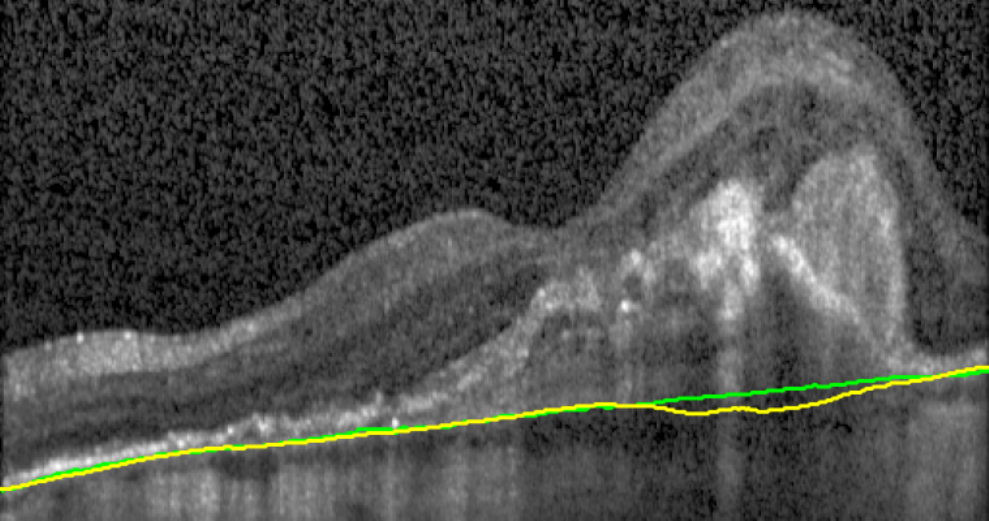}
    \vspace{0.01cm}
    \end{minipage}
    \hfill
    \begin{minipage}{.24\textwidth}
        \centering
        \includegraphics[width=1.0\textwidth]{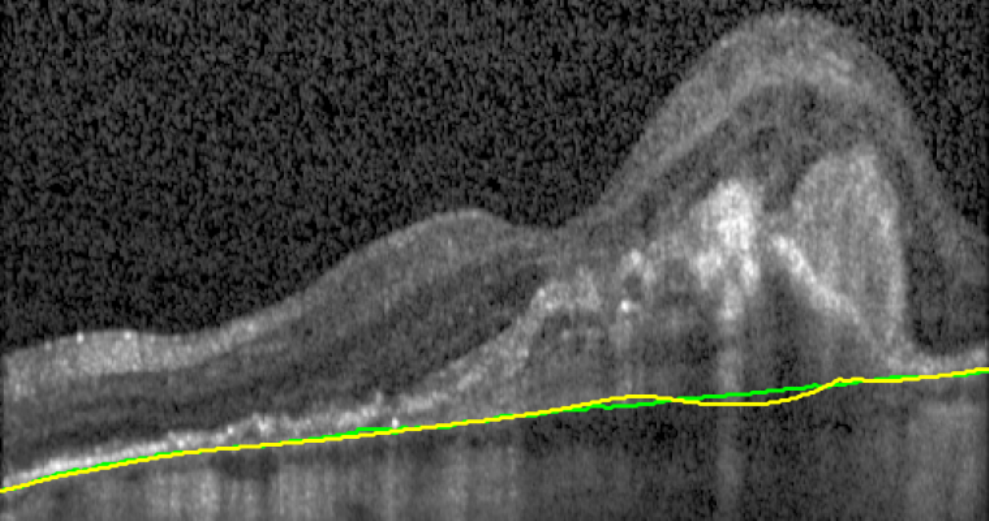}
    \vspace{0.01cm}
    \end{minipage}
    \begin{minipage}{.24\textwidth}
    \centering
        \includegraphics[width=1.0\textwidth]{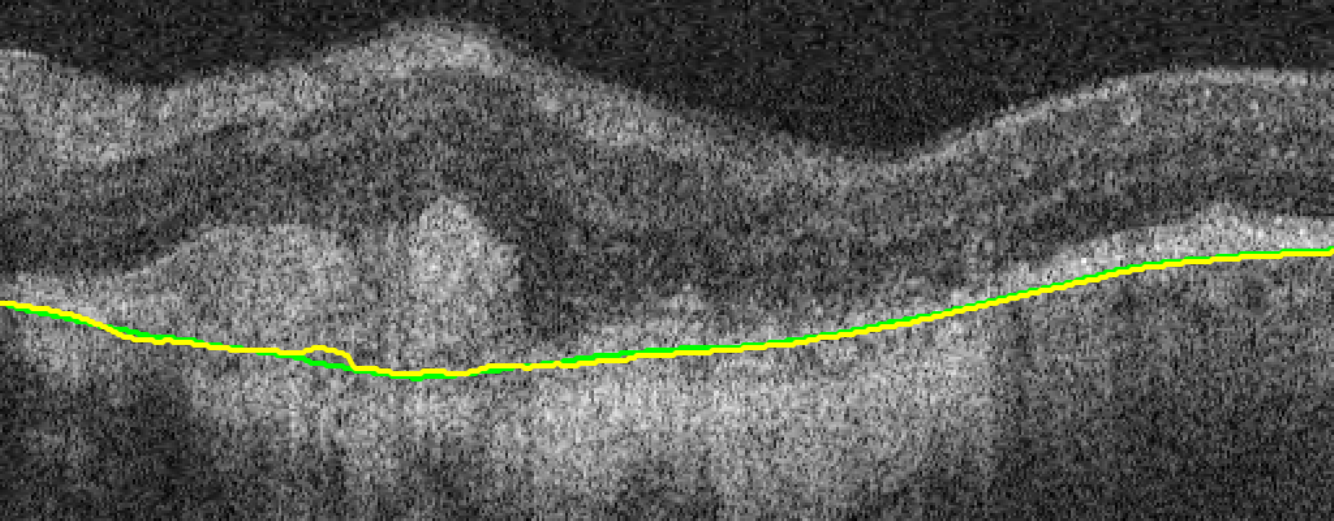}
        \caption*{(a) ReLayNet}
    \end{minipage}
    \hfill
    \begin{minipage}{.24\textwidth}
    \centering
        \includegraphics[width=1.0\textwidth]{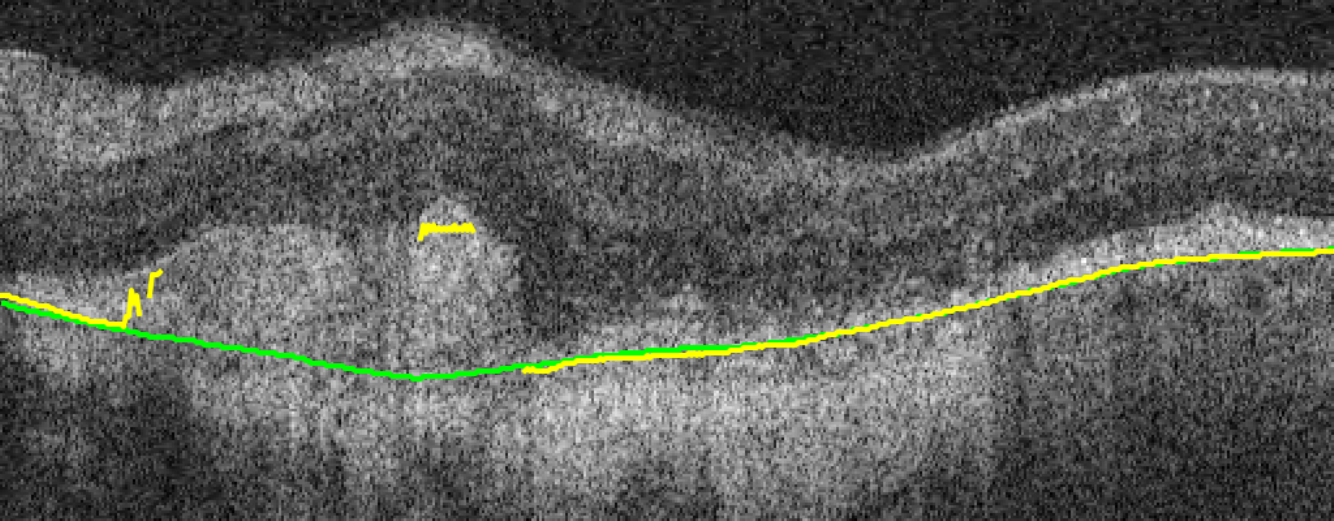}
        \caption*{(a) DexiNed \label{fig:qualitative_results_dexined}}
    \end{minipage}
    \hfill
    \begin{minipage}{.24\textwidth}
        \centering
        \includegraphics[width=1.0\textwidth]{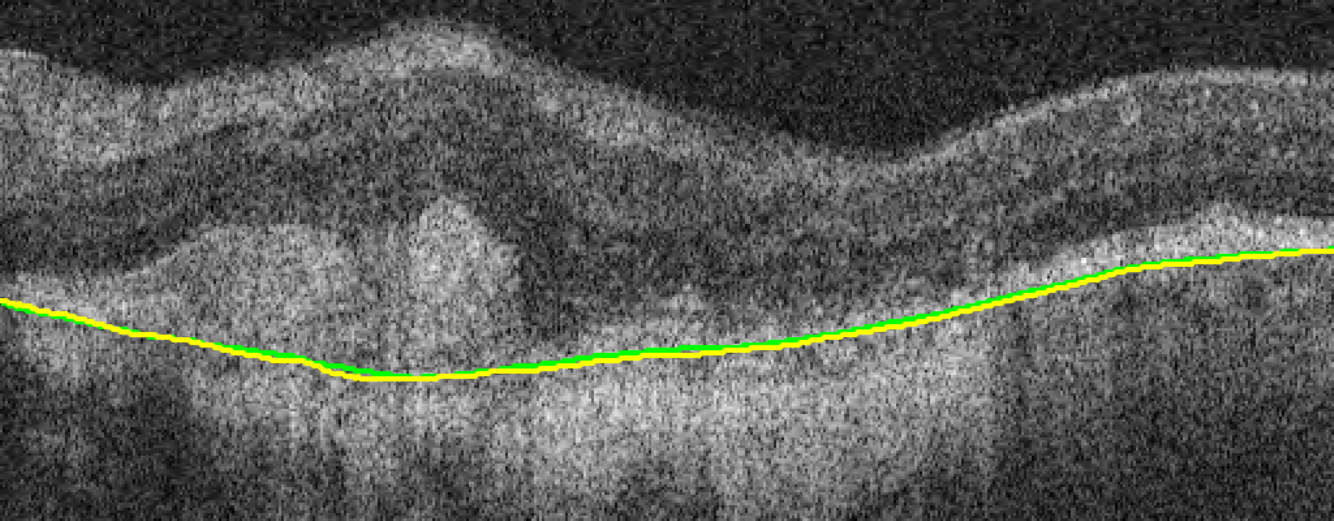}
        \caption*{(b) BFC-DN}
    \end{minipage}
    \hfill
    \begin{minipage}{.24\textwidth}
        \centering
        \includegraphics[width=1.0\textwidth]{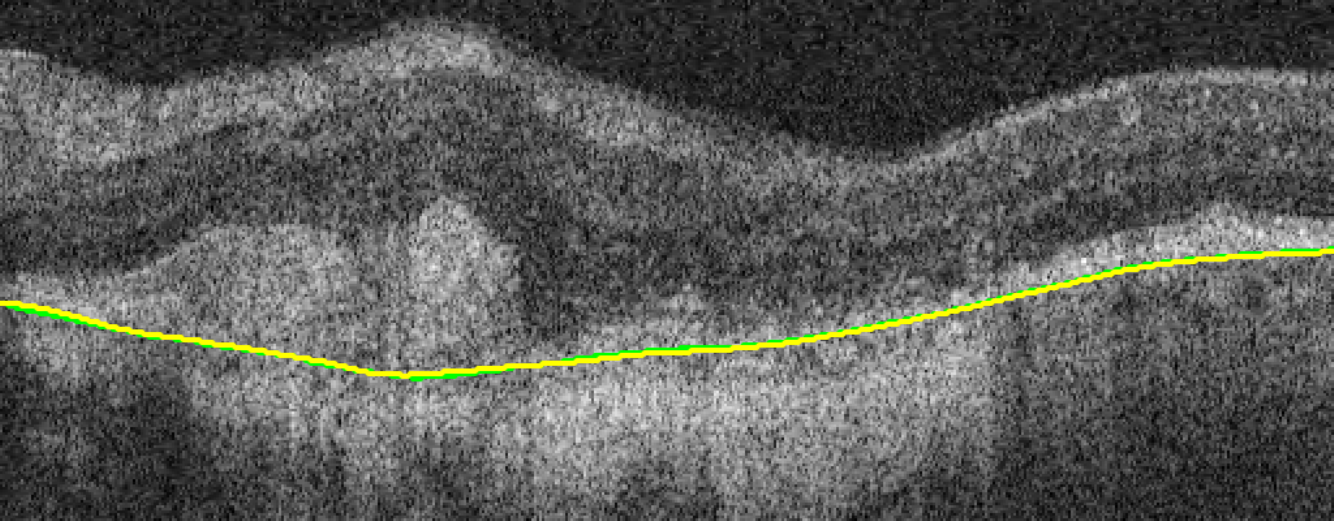}
        \caption*{\NVN{(c) Proposed}}
    \end{minipage}
    \caption{Sample segmentation results (yellow) from ReLayNet (a), DexiNed (b), BFC-DN (c), and our method (d), with the corresponding ground truth (green). Note that for DexiNed the segmentation result is not always present.}
    \label{fig:qualitative_results}
\end{figure*}

The performance of our approach in comparison with other state-of-the-art and baseline methods is shown quantitatively and qualitatively in Table~\ref{tab:our_results} and Fig.~\ref{fig:qualitative_results}, respectively.
\NVN{In terms of baseline performance, Table~\ref{tab:our_results} shows that \emph{Proposed w/o $\mathbf{\mathcal{L}_3}$, w/o TPS} is superior to the 2D surface prediction methods ReLayNet \cite{2017_Roy} and BFC-DN \cite{2018_Sedai_CONF}, and to the multi-step edge-detection method DexiNed \cite{2021_Sousa}. }
\NV{For DexiNed \cite{2021_Sousa} in particular, there are cases where the segmentation fails to provide any positions at certain A-scans (Fig.~\ref{fig:qualitative_results} third and fourth row). With this in mind, for quantitative evaluation, these A-scans were ignored exclusively for DexiNed and thus the reported values for this model are to be understood as a lower boundary of the segmentation error.}

\NV{While similar to ReLayNet \cite{2017_Roy}, being a pixel-wise segmentation method as well, the BFC-DN \cite{2018_Sedai_CONF} model resulted in a much more robust segmentation, although it still did not always produce smooth surfaces, especially in noisy scans (Figure~\ref{fig:qualitative_results}, second row).
}

\begin{figure}[tb]
    \begin{minipage}{.49\textwidth}
        \centering
        \includegraphics[width=1.0\textwidth]{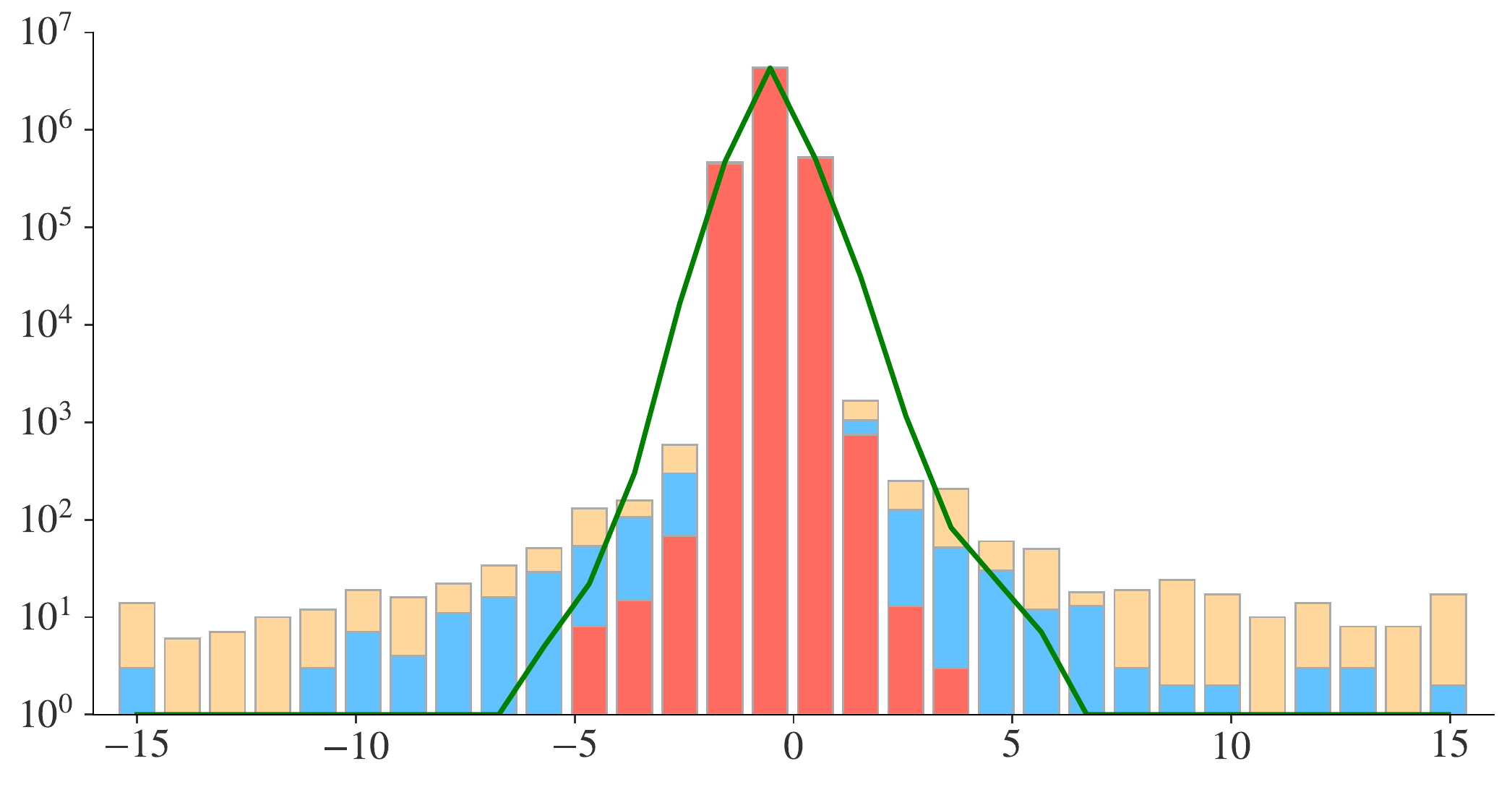}
        \caption*{(a)}
        \includegraphics[width=1.0\textwidth]{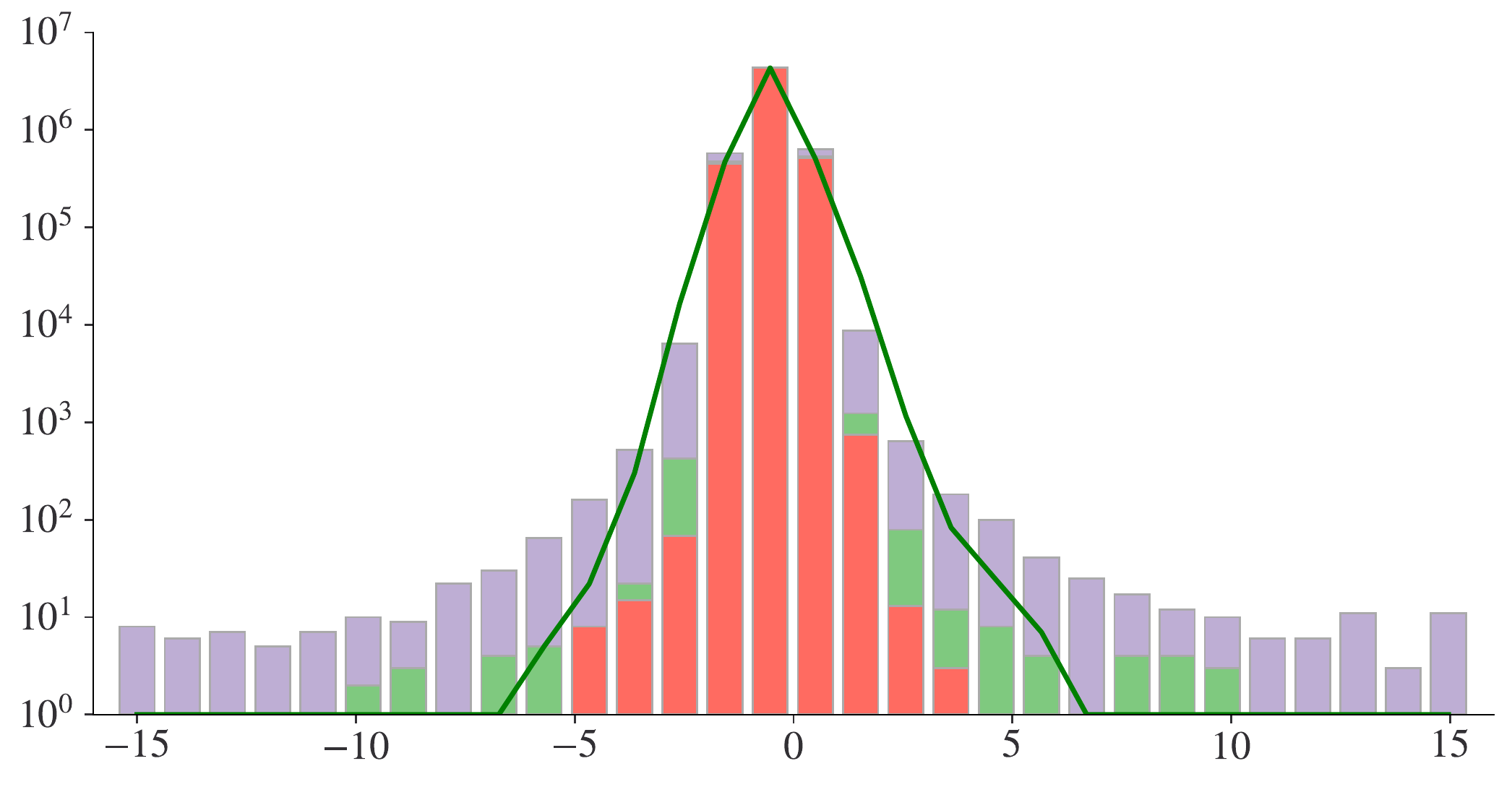}
        \caption*{(b)}
    \end{minipage}
    \caption{\NVN{Histogram of the distances between the BM positions in adjacent columns for the reference standard (green line), \textbf{(a)} \emph{ReLayNet} (orange bars),  \emph{Proposed w/o $\mathbf{\mathcal{L}_3}$, w/o TPS} (blue bars) and \emph{Proposed w/o TPS} (red  bars), \textbf{(b)} \emph{DexiNed} (purple bars) and \emph{BFC-DN} (light green bars) in our internal dataset. The units are in \emph{px}, where 1px $\approx 3.9~\upmu$m.}}
    \label{fig:our_curvature}
\end{figure}

\NV{\NVN{The ablation study (Table~\ref{tab:our_results}) indicates a significant improvement in all of the evaluation metrics (Wilcoxon $p < 0.01$) when using the curvature term $\mathcal{L}_3$ in addition to $\mathcal{L}_1 + \mathcal{L}_2$}.} The larger reduction in RMSE shows that the proposed loss term is successful in dampening the large outliers. The post-processing step with TPS further reduces the error compared to the reference standard (Wilcoxon $p < 0.05$). Due to the introduction of the curvature-based loss term the number of A-scans having a distance larger than 15$\upmu$m was reduced to 0 and in general a smoother surface can be observed (Figure~\ref{fig:our_curvature}). Ultimately, our method produces a more accurate segmentation than other state-of-the-art methods.

\begin{figure}[tb]
        \centering
        \includegraphics[width=0.45\textwidth]{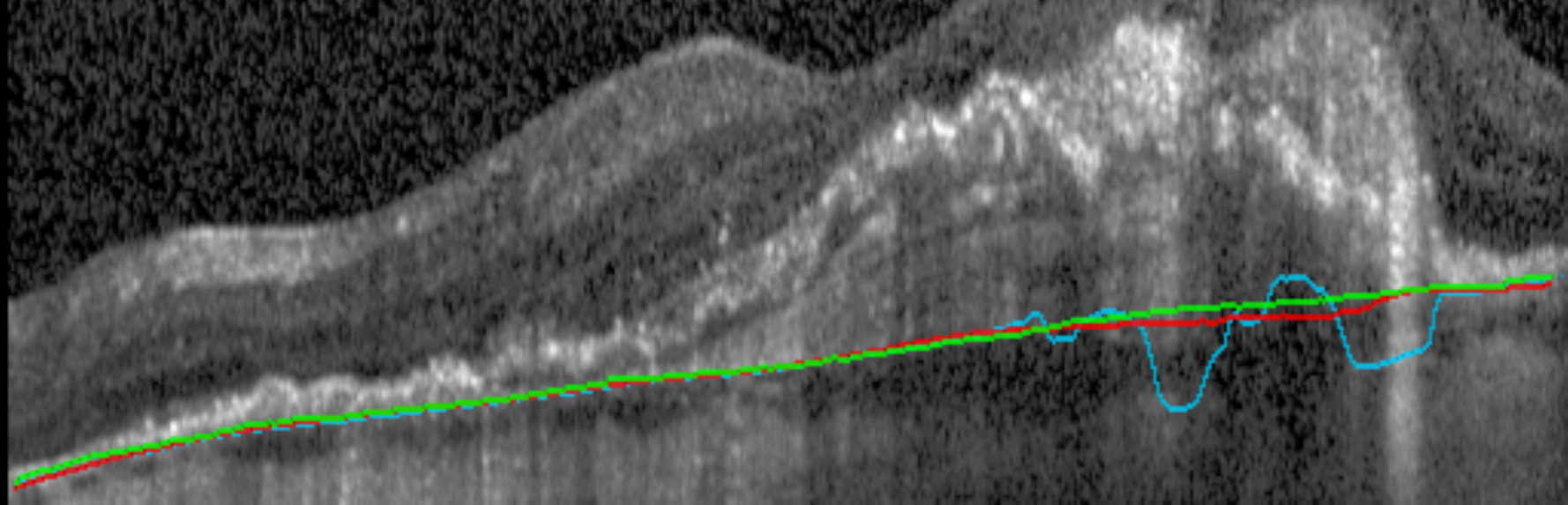}\vspace{1pt}
        \includegraphics[width=0.45\textwidth]{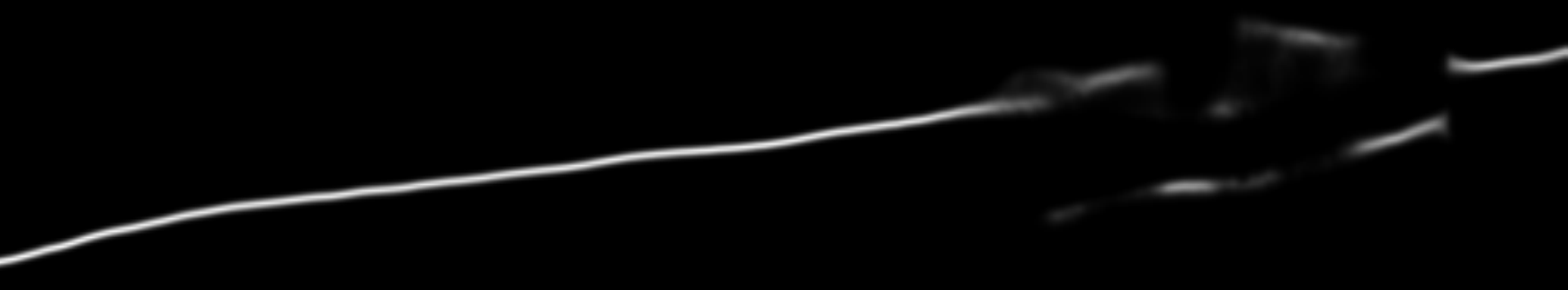}\vspace{1pt}
        \includegraphics[width=0.45\textwidth]{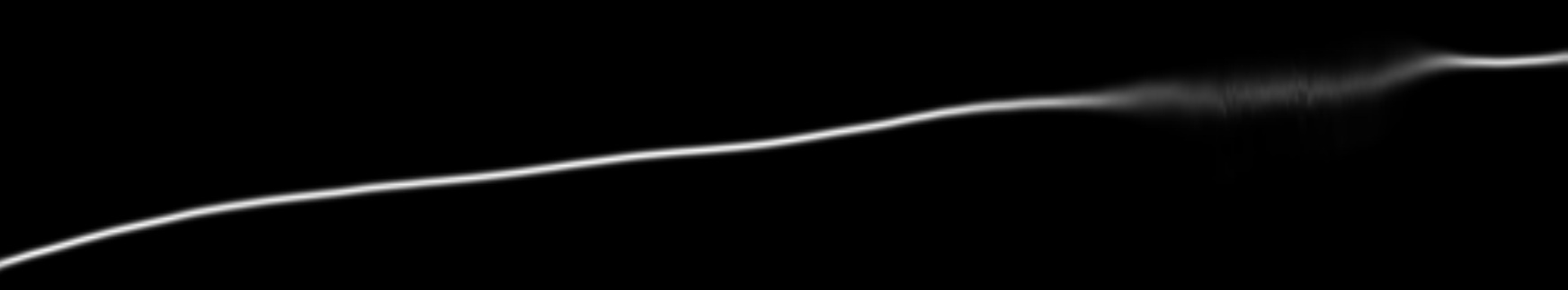}\vspace{1pt}
        \includegraphics[width=0.45\textwidth]{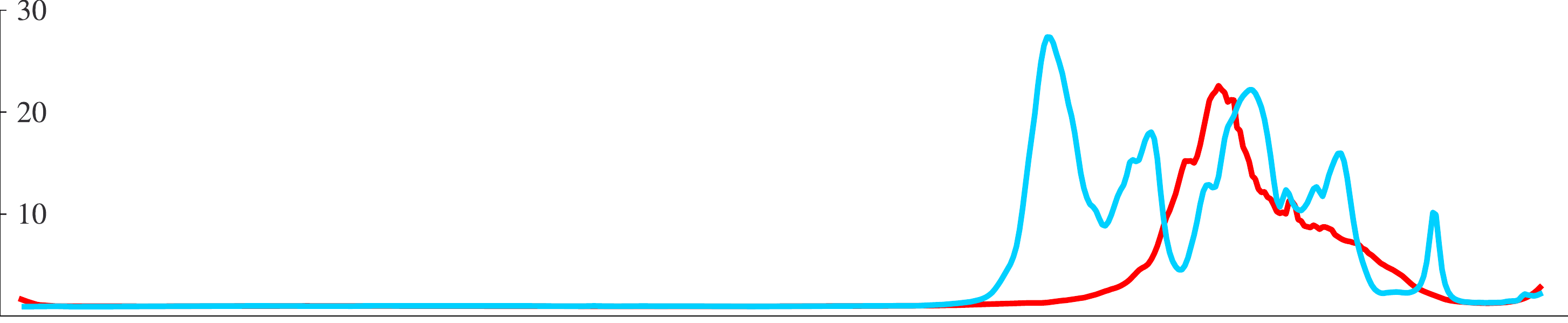}

    \caption{Representative example of the influence of the curvature preserving loss term on segmentation performance for a region of a B-scan. (Top) The relevant crop of the B-scan, with the reference standard (green), the prediction without the curvature term (blue), and the prediction with the curvature term (red). In addition, a visualization of the associated A-scan output mass function of the segmentation without (Middle-Top) and with (Middle-Bottom) the curvature term. The calculated corresponding uncertainties are also shown. (Bottom)}
    \label{fig:curvature_vs_nocurvature}
\end{figure}

A representative example of the influence of the curvature term $\mathcal L_3$ is shown in Fig.~\ref{fig:curvature_vs_nocurvature}. \NVN{The proposed method with the help of the curvature term correctly finds the BM as opposed to \textit{Proposed w/o $\mathbf{\mathcal{L}_3}$ w/o TPS}}. The associated uncertainty is also adequately higher, signaling the lower confidence of the BM position estimate in this region. In contrast, the absence of $\mathcal L_3$ leads to a poor uncertainty estimation in some incorrect segmentation regions.

\begin{table}[t]
\centering
    \begin{tabular}{|l||c|c|c|}
    \hline
    \textbf{Model}           & \textbf{MAE} [$\upmu$m]                     &  \textbf{RMSE} [$\upmu$m]                     \\ \hhline{|=||=|=|=|}

    \textbf{ReLayNet \url{~} \cite{2017_Roy}}       & $5.73 \pm 1.66 $               & $11.40 \pm  10.11$         \\ \hline
    \textbf{DexiNed \url{~} \cite{2021_Sousa}}       & $5.04^{*} \pm 1.63 $               & $7.53^{*} \pm  13.20$         \\ \hline
    \textbf{BFC-DN  \cite{2018_Sedai_CONF} }       & $4.63^{*} \pm 1.62 $               & $6.58^{*} \pm  5.57$         \\ \hline
    \textbf{\NVN{Proposed w/o $\mathbf{\mathcal{L}_3}$, w/o TPS \url{~} \cite{2021_He}}}            & $4.29^{*} \pm 1.73    $          & $6.34^{*} \pm 7.00$           \\ \hline
    \textbf{\NVN{Proposed w/o TPS}}   & $4.11^{*} \pm 1.64 $                & $5.94^{*} \pm 5.36$       \\ \hline
    \textbf{\NVN{Proposed}} & $\mathbf{ 4.10 \pm 1.63}$    & $\mathbf{ 5.88^{*} \pm 5.13}$  \\ \hline
    \end{tabular}%
    \caption{\NVN{Performance of the baseline methods (DexiNed, ReLayNet, BFC-DN) and influence of the curvature preserving loss term $\mathcal{L}_3$ and thin plane splines post-processing (TPS) in the proposed architecture.} MAE: mean absolute error; RMSE: root mean square error. Asterisk indicates Wilcoxon signed-rank test $p$-value $<$ 0.05 compared to the model in the previous row. \NV{Note, that the missing values in the DexiNed segmentations were left out from the calculation.}}
    \label{tab:our_results}
\end{table}

\subsection{Performance across AMD stages and OCT devices}
\begin{figure}[b]
        \centering
        \includegraphics[width=0.45\textwidth]{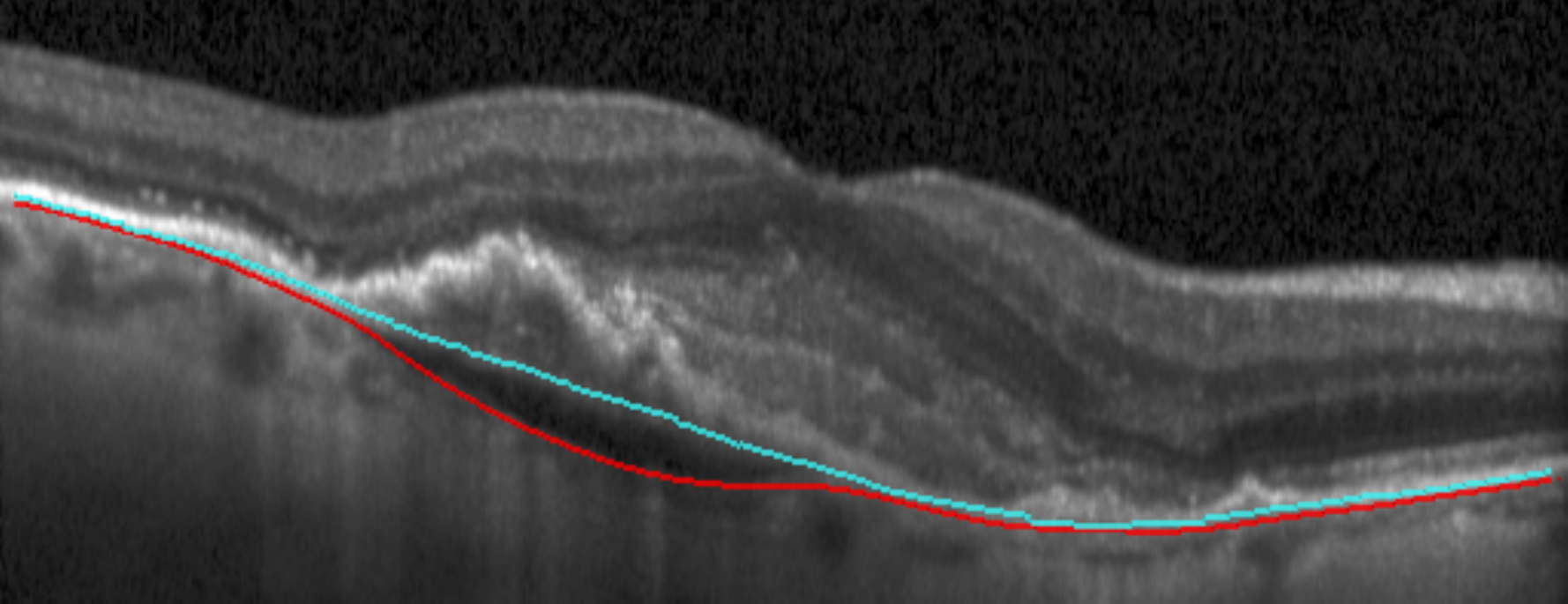}
    \caption{An example from our internal dataset where a cleft was correctly captured by our method with $\mathcal{L}_3$ (red), while not without $\mathcal{L}_3$ (blue).}
    \label{fig:ours_bad_groundtruth}
\end{figure}

The models' performance for a particular AMD stage and OCT vendor combination is detailed in Table~\ref{tab:disease_vendor}. In general, the introduction of the curvature term improves on the segmentation results across the disease stages and device vendors. \NVN{The proposed model was able to correctly handle more complicated cases, including clefts\cite{2018_Kim} (Fig.~\ref{fig:ours_bad_groundtruth}), which were not identified by \emph{Proposed w/o $\mathcal{L}_3$}  w/o TPS}. We hypothesise that the reason for this is that the unusual curvature of the BM under a cleft leads to a higher $\mathcal{L}_3$ value during the training and thus the model is able to learn these cases, despite their relative scarcity. The TPS-based post-processing further improves the position regression. However, it resulted in a slight performance drop in nAMD Cirrus scans, which was largely due to some severely misplaced B-scans with bad image quality, showing the limitations of the current heuristic to align the scans for the 3D interpolation. In general, the model performed 9\% better on Spectralis scans than on Cirrus scans. Possible contributing factors were the skew in the training set, since it contained more than twice as many Spectralis scans as Cirrus scans, and the initial resizing of the input images, which yielded a better effective digital axial resolution of 3.75$~\upmu$m in Spectralis scans while 3.91~$\upmu$m in Cirrus scans.

\begin{table*}[htb]
    \centering
\begin{tabular}{|l|l||c|c|c|}
\hline
\multicolumn{2}{|l||}{\diagbox[width=4cm]{\textbf{Stage/Vendor}}{\textbf{Method}}}               & \NVN{\textbf{Proposed w/o $\mathbf{\mathcal{L}_3}$, w/o TPS \url{~} \cite{2021_He}}} & \textbf{\NVN{Proposed w/o TPS}} & \textbf{\NVN{Proposed}}\\  \hhline{|==||=|=|=|}
\multirow{2}{*}{\textbf{iAMD}} & \textit{Cirrus}     & N/A           & N/A                & N/A                    \\ \cline{2-5} 
                               & \textit{Spectralis} & $5.70 \pm 4.75 $  & $5.38 \pm 3.30 $             & $ \mathbf{ 5.34 \pm 3.38 }$         \\ \hline
\multirow{2}{*}{\textbf{nAMD}} & \textit{Cirrus}     & $7.50 \pm 7.34 $  & $\mathbf{7.02 \pm 6.65 }$             & $7.14 \pm 6.69 $                  \\ \cline{2-5} 
                               & \textit{Spectralis} & $6.11 \pm 5.18$  & $5.94 \pm 4.73 $             & $\mathbf{5.86 \pm 4.49}$         \\ \hline
\multirow{2}{*}{\textbf{GA}}   & \textit{Cirrus}     & $6.11 \pm 6.43 $  & $5.57 \pm 5.48 $             & $\mathbf{5.39 \pm 5.02 }$         \\ \cline{2-5} 
                               & \textit{Spectralis} & $7.93 \pm 9.87 $  & $6.86 \pm 6.49 $             & $\mathbf{6.53 \pm 5.34 }$         \\ \hline
\multirow{2}{*}{\textbf{All}}  & \textit{Cirrus}     & $6.74 \pm 6.90 $  & $6.24 \pm 6.12$      & $\mathbf{6.21 \pm 6.02}$                  \\ \cline{2-5} 
                               & \textit{Spectralis} & $6.28 \pm 7.08 $  & $5.79 \pm 4.83 $               & $\mathbf{5.67 \pm 4.25 }$         \\ \hline
\end{tabular}
\caption{The RMSE (mean $\pm$ std) of the ablation studies in $\upmu$m \hl{across} different device vendors and AMD stages.}
\label{tab:disease_vendor}
\end{table*}

\subsection{Validation on an external test set under domain shift}

Table~\ref{tab:duke_dataset} shows a comparison of our method with state-of-the-art methods reported on the public Duke dataset. The errors are reported in pixel units, to match what has been published previously, but note that the values can be transformed into $\upmu$m unit by multiplying with $3.24$, the digital axial resolution of a Bioptigen OCT device \cite{2014_Farsiu}.

On this dataset our method yields a better or equal MAE as the current state-of-art result of Sousa\etal \cite{2021_Sousa} (DexiNed), albeit, with a larger standard deviation. It should be noted that the results are not directly comparable: \NV{We tested our method on the entire Duke dataset, in contrast to a reduced subset used in~\cite{2021_Sousa}, as 1) we did not train with any Duke data and 2) the test splits of Sousa\etal\cite{2021_Sousa} and Sedai\etal\cite{2018_Sedai_CONF} were not publicly available.} Our method was also trained on a larger training set, but with scans from a distinct population and different OCT devices. The results clearly show that our method generalizes well to an external dataset and under an image domain shift caused by a different OCT vendor.

Of note, we identified a few wrong reference annotations in the dataset, with selected B-scans illustrated in Fig.~\ref{fig:duke_bad_groundtruth}. This lowers the upper limit on the achievable segmentation performance, as well as it influences the uncertainty estimation evaluation. However, since they were found only in small subset of the whole dataset, a comparison between the different methods on this dataset is still meaningful.

\begin{table}[bt]
\centering
\begin{tabular}{|c||c|c|c|}\hline
                                  & \textbf{Method} & \textbf{MAE} & \textbf{Std} \\ \hhline{|=||=|=|=|}
\multirow{2}{*}{\textbf{AMD}}     & DexiNed \cite{2021_Sousa}        & 0.70         & \textbf{0.13}         \\ \cline{2-4} 
                                  & \textbf{\NVN{Proposed}}            & 0.70         & 0.24         \\ \hline\hline
\multirow{2}{*}{\textbf{Control}} & DexiNed \cite{2021_Sousa}        & 0.59         & \textbf{0.08}         \\ \cline{2-4} 
                                  & \textbf{\NVN{Proposed}}   & \textbf{0.47}         & 0.11         \\ \hline\hline
\multirow{8}{*}{\textbf{All}}     & RNN-GS \cite{2018_Kugelman}         & 2.07         & 4.31         \\ \cline{2-4} 
                                  & CNN-GS \cite{2018_Kugelman}         & 2.31         & 4.60         \\ \cline{2-4} 
                                  & FCN-GS \cite{2018_Kugelman}         & 1.53         & 3.50         \\ \cline{2-4} 
                                  & CapsNet \cite{2019_Santos_DISSERTATION}        & 1.09         & 2.49         \\ \cline{2-4} 
                                  & DeepForest \cite{2019_Chen}     & 1.24         & 0.52         \\ \cline{2-4} 
                                  & WAVE  \cite{2020_Lou}          & 1.90         & 3.01         \\ \cline{2-4} 
                                  & BFC-DN  \cite{2018_Sedai_CONF}          & 0.97         & 0.86         \\ \cline{2-4} 
                                  & DexiNed \cite{2021_Sousa}      & 0.66         & \textbf{0.12}         \\ \cline{2-4} 
                                  & \textbf{\NVN{Proposed}}   & \textbf{0.63}         & 0.23         \\ \hline
\end{tabular}
\caption{Comparison with related works on the Duke dataset with respect to MAE and its standard deviation (Std.)}
\label{tab:duke_dataset}
\end{table}

\subsection{Uncertainty estimation}

\begin{figure}[tb]%
    \centering
    \subfloat[\centering Our dataset]{{\includegraphics[width=0.40\textwidth]{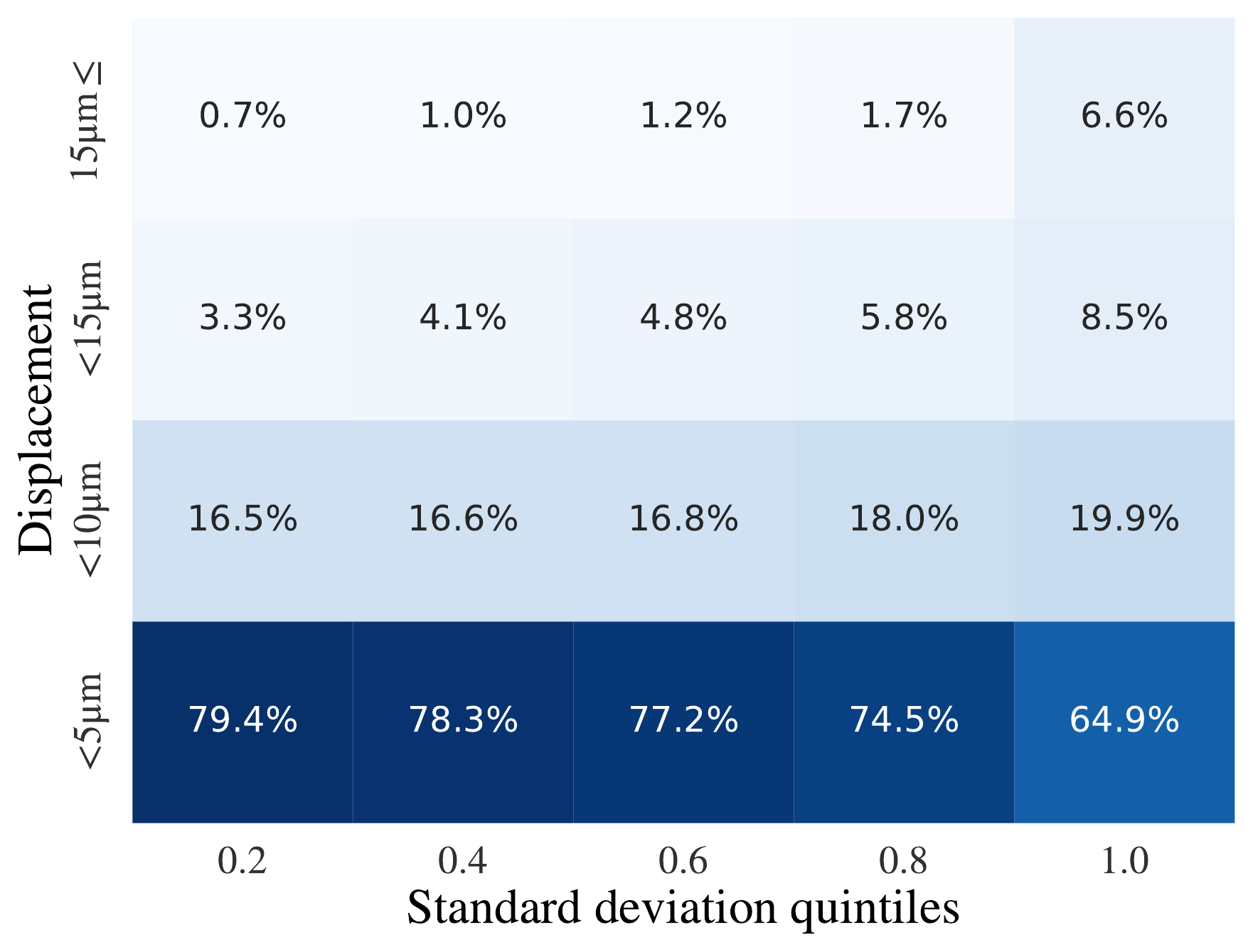} }}%
    \qquad
    \subfloat[\centering Duke]{{\includegraphics[width=0.40\textwidth]{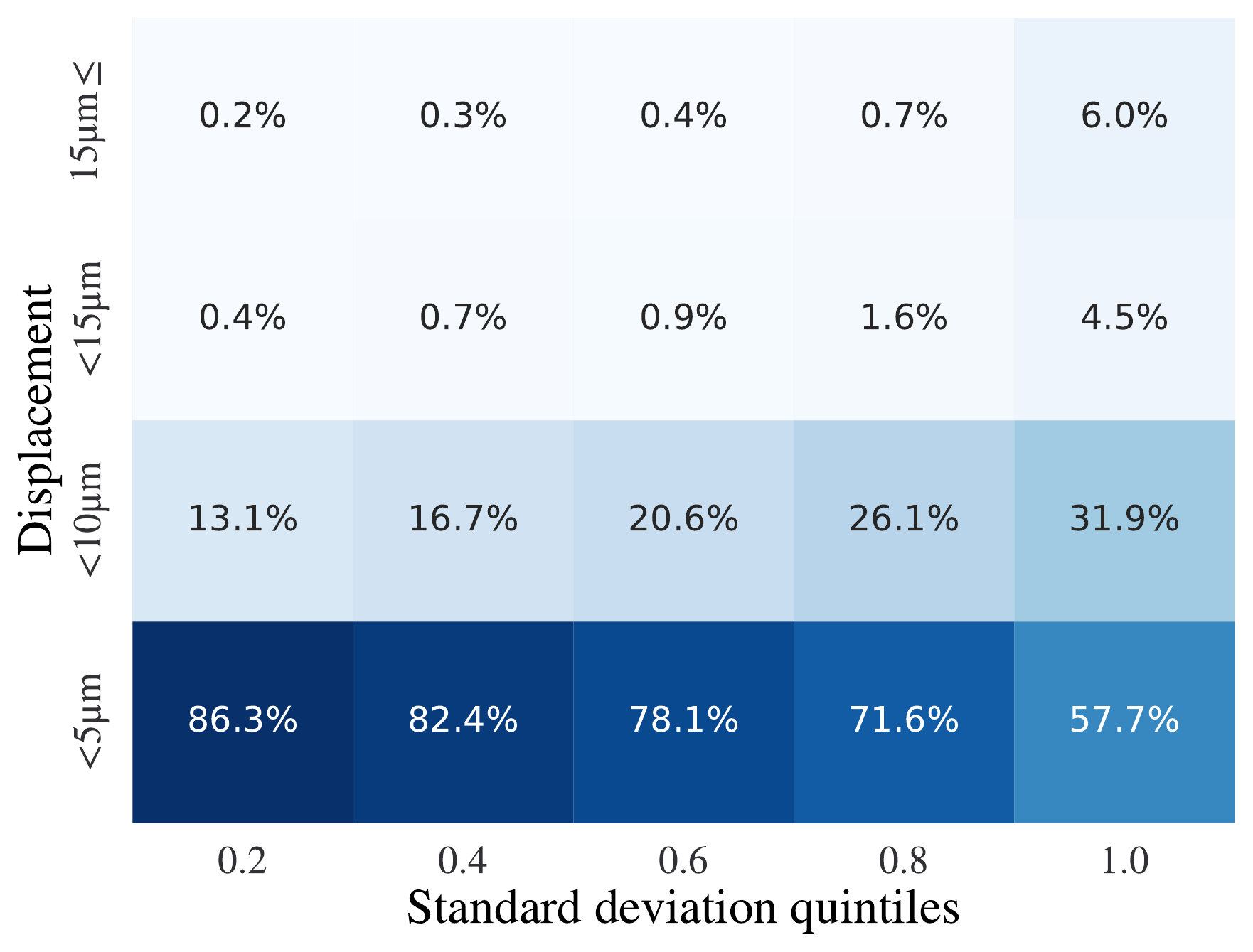} }}%
    \caption{The distribution of the displacement of the predicted BM positions against the quintiles of the predicted uncertainty. The lower is the uncertainty, the more accurate is the prediction.}%
    \label{fig:ascan_vs_std}%
\end{figure}

The mean standard deviation (corresponding to the uncertainty) at a volume level was found to be positively correlated with the MAE (SCC = 0.74, $p < 0.001$). On the B-scan level this correlation was observed to be weaker (SCC = 0.51, $p < 0.001$) and even more so on A-scan level (SCC = 0.22, $p < 0.001$). The low correlation coefficient on A-scan level results from the conservative nature of the uncertainty estimation, as in a number of times the model is able to find the correct BM position, however, with a high uncertainty (Fig.~\ref{fig:ascan_vs_std}). Although the correlation between the A-scan displacement and the uncertainty measure is not high, the improvement resulting from the TPS interpolation shows that even the A-scan-level uncertainties can be used for correcting the possibly wrong position regressions.

Fig.~\ref{fig:ascan_vs_std} provides an overview of the distribution of segmentation errors by the standard deviation quintiles, showing a strong relation between the uncertainty measure and the displacement error. The lower the estimated uncertainty is, the lower is the share of A-scans with a displacement below 15$\upmu$m, and A-scans with higher uncertainties tend to have larger displacements.

\begin{figure}[tb]%
    \centering
    \subfloat[\centering Our dataset]{{\includegraphics[width=0.30\textwidth]{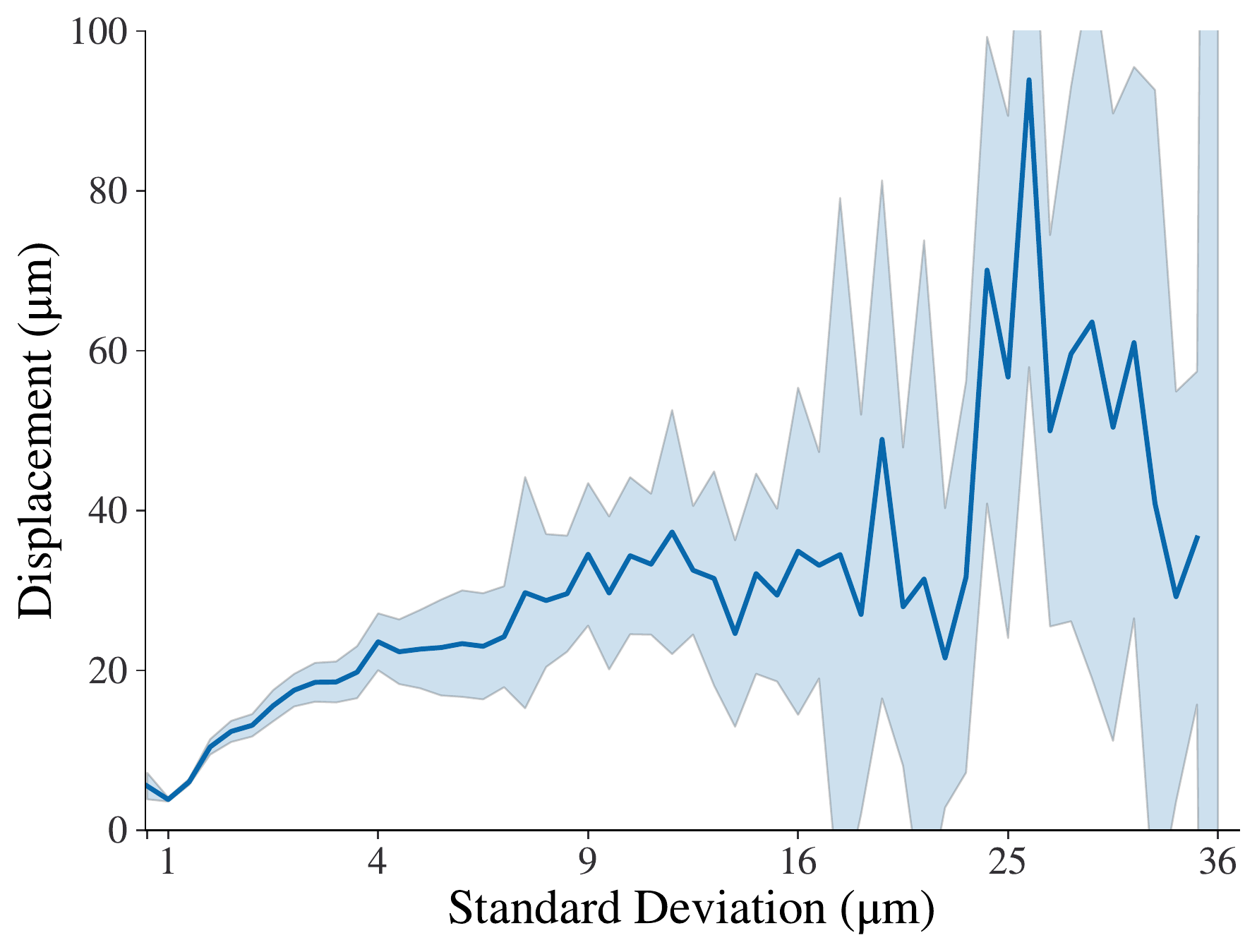} }}%
    \qquad
    \subfloat[\centering Duke]{{\includegraphics[width=0.30\textwidth]{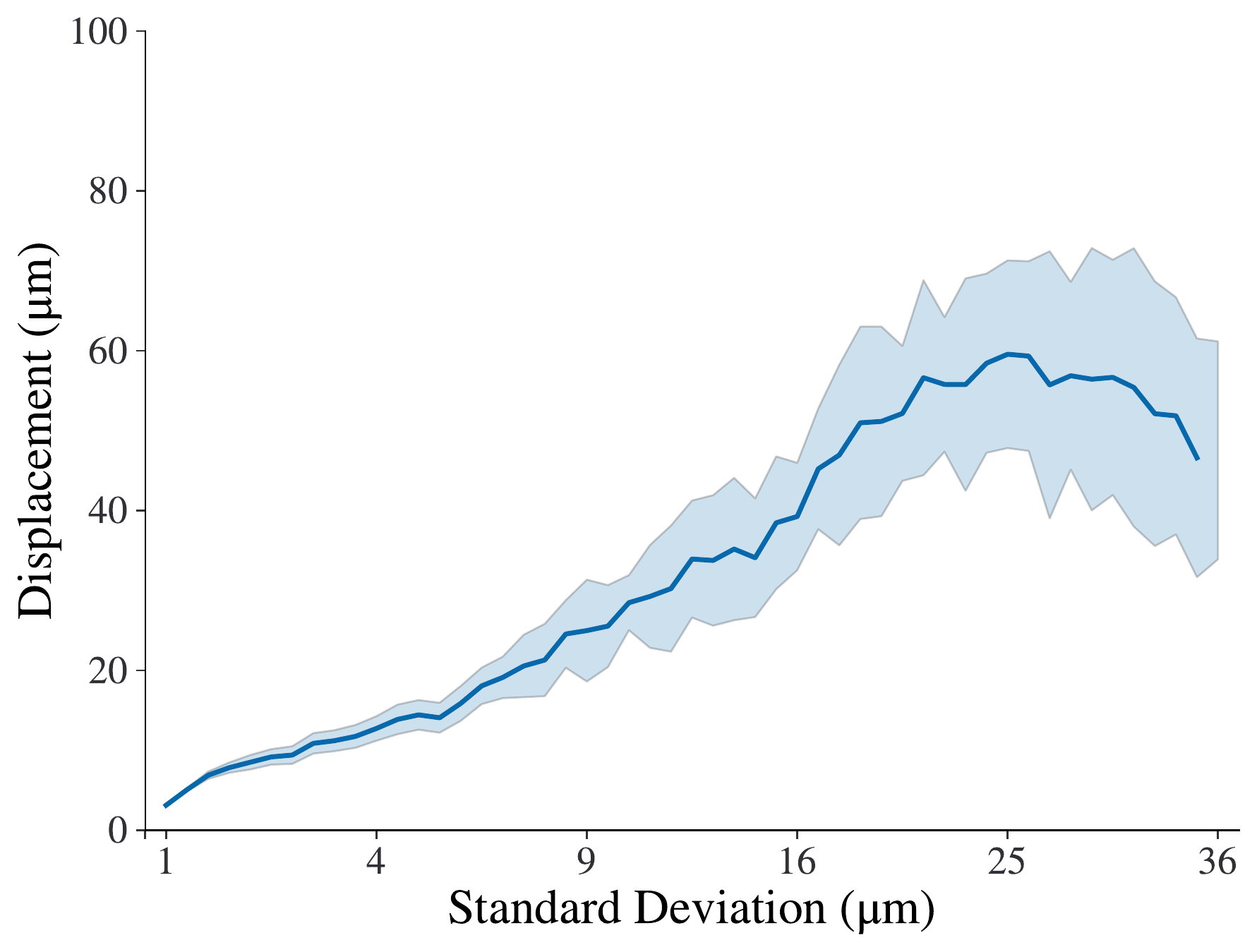} }}%
    \caption{The relation between the mean displacement and the standard deviation rounded to one decimal place. The shaded areas represent the 95\% confidence interval of the mean.}%
    \label{fig:avg_error_vs_std}%
\end{figure}

The clear relationship could be found between the average A-scan-wise displacement and the standard deviation values in our internal dataset and on Duke (Fig.~\ref{fig:avg_error_vs_std}). For A-scans with small uncertainty values, where the majority of them lies, the uncertainty value was well-calibrated with the displacement. However, for larger uncertainty values, the connection was less clear partly due to large confidence intervals as only 0.01\% of the A-scans had a standard deviation above 5.0$\upmu$m.

\begin{figure}[tb]
        \centering
        \includegraphics[width=0.45\textwidth]{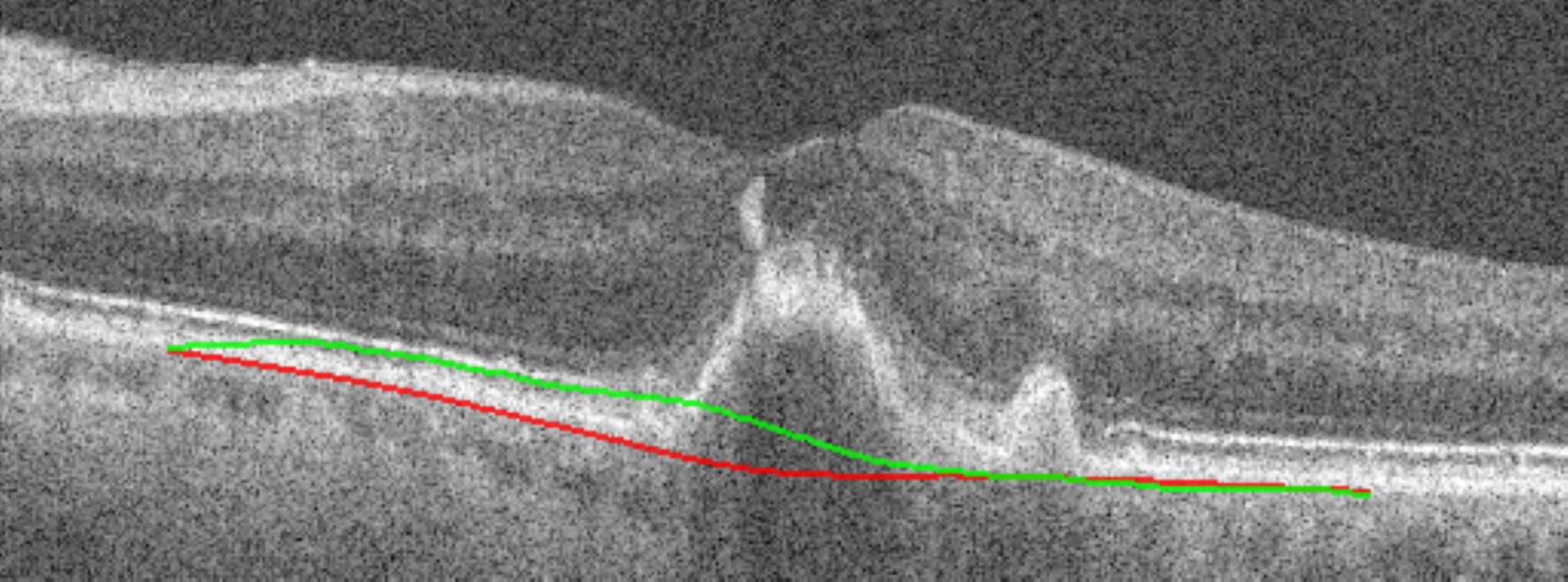}\vspace{1pt}
        \includegraphics[width=0.45\textwidth]{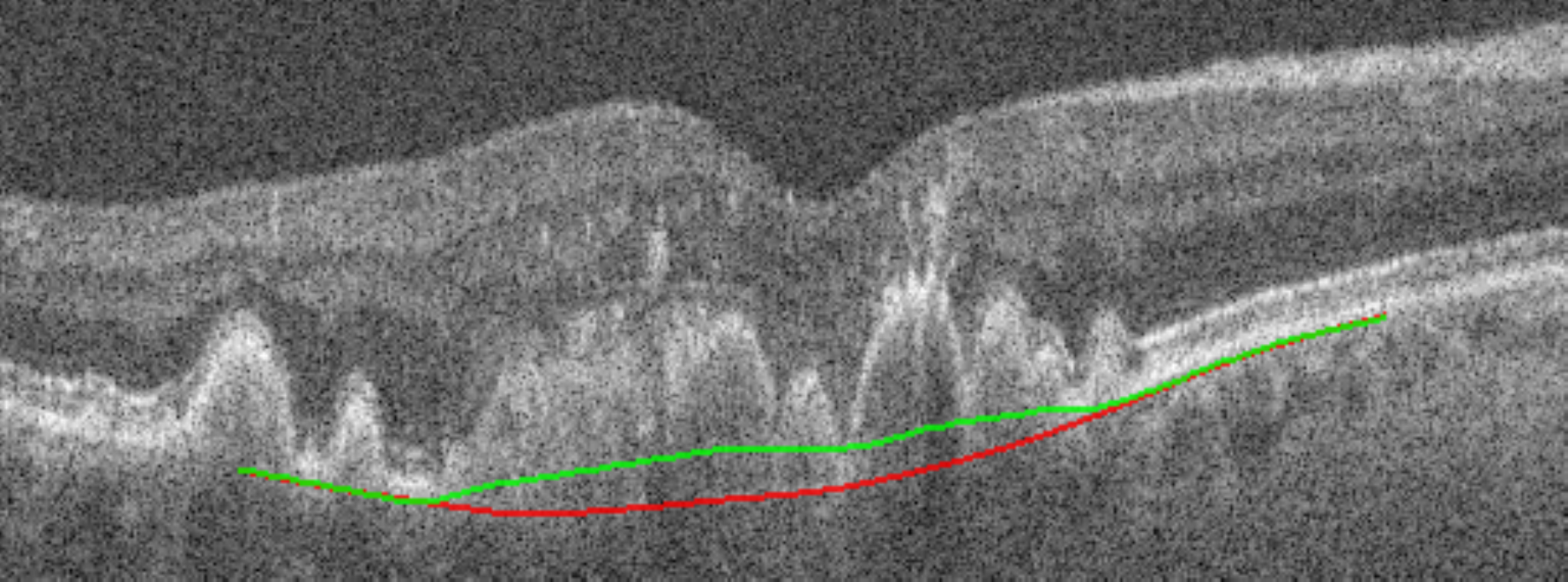}

    \caption{Sample B-scans from the Duke dataset with wrong BM reference standard (green), but correct segmentation (red). 
    }
    \label{fig:duke_bad_groundtruth}
\end{figure}

\begin{figure}[tb]
    \centering
    \includegraphics[width=0.45\textwidth]{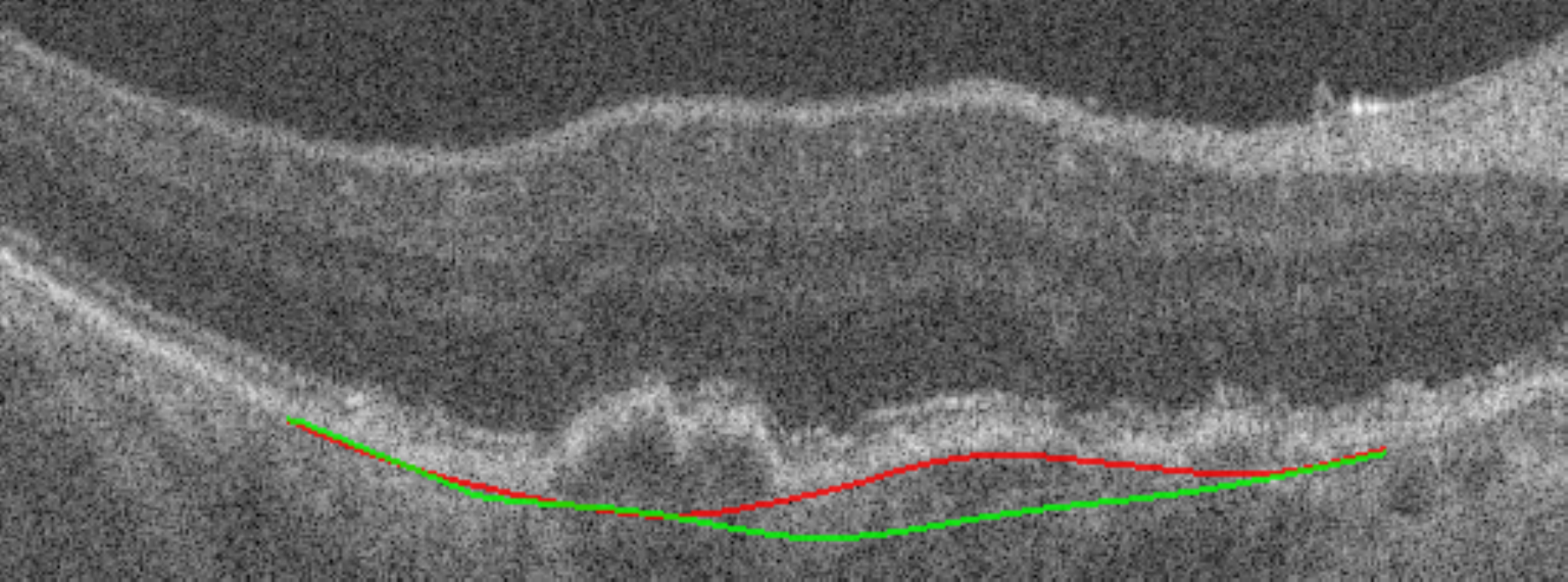}\vspace{1pt}
    \includegraphics[width=0.45\textwidth]{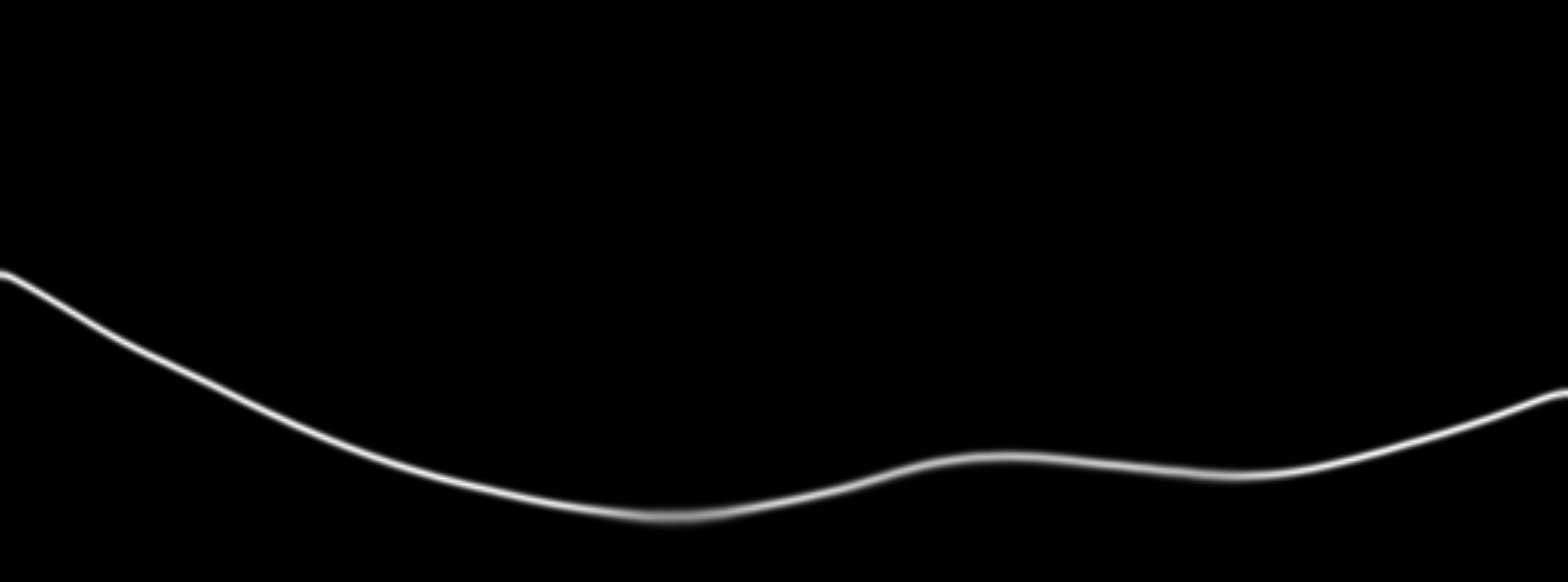}
    \caption{The segmentation result from a B-scan from the single scan in the Duke database, in which the proposed method produced a wrong estimate (red) with low uncertainty. The reference standard is shown as green.}
    \label{fig:bad_segmentation_high_certainty}
\end{figure}

We visually inspected the B-scans containing A-scans with segmentation errors larger than 15$\upmu$m and low uncertainty ($\sigma < 1.0$). The majority of such large error margins were caused by misplacements in the reference standards, as exemplified in Fig.~\ref{fig:duke_bad_groundtruth}. However, we also found one B-scan in the Duke dataset, in which our method resulted in wrong BM positions with a low uncertainty (Fig.~\ref{fig:bad_segmentation_high_certainty}).

\section{Discussion}
\label{sec:discussion}

\NV{The paper proposed a deep learning method that provides an anatomically coherent segmentation of the BM in an accurate, reliable and robust manner, regardless of the different retinal morphologies and acquisition settings. In particular, we showed that using a curvature preservation term in the loss function allowed obtaining smoother segmentation and improved its overall accuracy over the current methods. At the same time, it helped to overcome the most common problem with the state-of-the-art automated segmentation models, the discontinuous or anatomically implausible surface position regressions.} The new term encodes a domain knowledge-based inductive bias that BM is a smooth surface, allowing more data-efficient learning and better generalization. Importantly, the method also measures the confidence of its output, which allows identifying and correcting the sections where the segmentation likely failed. \NV{Lastly, to the best of our knowledge, this is the first work where the proposed method underwent large-scale validation on scans covering all three stages of AMD, with different acquisition settings on both an internal dataset and well as on an external test set, thus more accurately representing its performance in a typical real-world clinical setting.}

Management of AMD, the leading cause of blindness in the developed world, would strongly benefit from  AI-based clinical decision support systems. Currently, the management is largely based on a \emph{qualitative} analysis of the retinal condition as captured by the OCT. Due to an increasingly ageing global population leading to an overwhelming amount of AMD patients, a higher-level of precision and automation will be required as well as a shift toward \emph{quantitative} OCT biomarkers. Such hallmark biomarkers in AMD are drusen volume (iAMD stage), PED volume (nAMD stage), and RPE thinning (GA stage), and they all rely on the ability to delineate BM in an automated, objective and repeatable manner. \NV{On the other hand, overreliance on automation carries inherent risks, as AI systems tend to provide overconfident predictions on both correct and incorrect results. Providing well-calibrated confidence scores gives an additional context for the interpretation of the results, and thus increases the trust in these systems and helps the clinical adoption of AI-based automated segmentation systems.}

The proposed method has a few limitations. First, the current surface prediction is largely based on A-scan properties and a limited neighboring context is being exploited. In fact, we conjecture that part of the performance improvement due to the curvature loss-term comes from allowing the network to utilize the global coherence of the segmentation, and to rely on neighboring regions to improve the quality of the result in the sections obscured by imaging artifacts or large anatomical abnormalities. Further improvement could be a reliance on a 3D U-Net \cite{2016_Cicek_CONF} to incorporate information from the adjacent B-scans. Second, a limited number of B-scans still showed low uncertainty, while segmenting the layer at the wrong position (Fig.~\ref{fig:bad_segmentation_high_certainty}). Uncertainty calibration of segmentation methods is an important and still open research question~\hl{\cite{2019_Jungo_CONF}}. 
Third, our pre-processing and post-processing steps are susceptible to eye motion artifacts. We could envision, as part of future work, that motion correction and segmentation can be trained to be performed simultaneously.

\NV{In summary, the proposed work advances the state of the art in achieving a robust and reliable layer segmentation with powerful generalization in the context of retinal OCT imaging. Its capability in providing anatomically coherent results, confidence estimates, as well as large-scale and out-of-sample validation, constitute important components of trustworthy AI systems, a necessity for integration of AI into clinical workflows~\hl{\cite{2021_GonzalezGonzalo}}.} Thus, we believe this work presents an important step toward the development of trustworthy AI tools for the management of patients with AMD.

\section*{Source Code}
The source code of the proposed method will be made publicly available upon acceptance at \url{https://github.com/ABotond/BM-Curvature}.

\section*{Funding}
The financial support by the the Christian Doppler Research Association, Austrian Federal Ministry for Digital and Economic Affairs, the National Foundation for Research, Technology and Development, and Heidelberg Engineering is gratefully acknowledged.

\section*{Disclosures}
The authors declare no conflicts of interest.

\bibliographystyle{IEEEtran}
\bibliography{bm-bib}

\end{document}